\RequirePackage{fix-cm}
\documentclass[twocolumn,epjc3]{svjour3}
\usepackage{pdfsync}
\usepackage{epsfig,epsf}
\usepackage{amsmath}
\usepackage{amsfonts}
\usepackage{amssymb}
\usepackage{dsfont}
\usepackage{slashed}
\usepackage{cancel}
\smartqed  
\DeclareMathOperator{\tr}{tr}
\DeclareMathOperator{\Li}{Li}

\newcommand \VEV [1] {\left\langle{#1}\right\rangle}
\newcommand \VVV [1] {\left\langle\left\langle{#1}\right\rangle\right\rangle}
\def\II{\hbox{{1}\kern-.25em\hbox{l}}}

\RequirePackage{graphicx}
\RequirePackage[numbers,sort&compress]{natbib}
\journalname{Eur. Phys. J. C}

\begin{document}

\title{Evolution equations beyond one loop from conformal symmetry
}


\author{V.~M.~Braun\thanksref{e1,addr1}
        \and
        A.~N.~Manashov\thanksref{e2,addr1,addr2} 
}

\thankstext{e1}{e-mail:  vladimir.braun@physik.uni-regensburg.de}
\thankstext{e2}{e-mail: alexander.manashov@physik.uni-regensburg.de}


\institute{Institut f\"ur Theoretische Physik, University of Regensburg, D-93040
Regensburg, Germany  \label{addr1}
           \and
         Department of Theoretical Physics,  St.-Petersburg State University,
199034 St.-Petersburg, Russia \label{addr2}
}

\date{Received: date / Accepted: date}

\maketitle

\begin{abstract}
We study implications of exact conformal invariance of scalar quantum field
theories at the critical point in non-integer dimensions for the evolution kernels of the
light-ray operators in physical (integer) dimensions.
We demonstrate that all constraints due the conformal symmetry  are encoded
in the form of the generators of the collinear $sl(2)$ subgroup.
Two of them, $S_-$ and $S_0$, can be fixed at all loops in terms of the evolution kernel,
while the generator of special conformal transformations, $S_+$, receives  nontrivial
corrections which can be calculated order by order in perturbation theory.
Provided that the generator $S_+$ is known at the $\ell-1$ loop order, one can fix the evolution
kernel in physical dimension to the $\ell$-loop accuracy up to terms that are invariant with respect to the tree-level generators.
The invariant parts can easily be restored from the anomalous dimensions.
The method is illustrated on two examples:
The $O(n)$-symmetric $\varphi^4$ theory in $d=4$ to the three-loop accuracy,
and the $su(n)$ matrix $\varphi^3$ theory in $d=6$ to the two-loop accuracy.
We expect that the same technique can be used in gauge theories e.g. in QCD.

\end{abstract}

\section{Introduction}
\label{intro}
It is well known that conformal symmetry of the QCD Lagrangian imposes strong constraints
on the leading-order (LO) correlation functions and operator renormalization, see
Ref.~\cite{Braun:2003rp} for a review. A schematic structure of the perturbation theory for a
generic quantity $\mathcal{Q}$ beyond the LO is usually conjectured to be
\begin{align}
  \mathcal{Q} =  \mathcal{Q}^{\rm con} + \frac{\beta(g)}{g}\Delta \mathcal{Q}
\label{conformal}
\end{align}
where $\mathcal{Q}$ is the result in the formal conformal limit, obtained by setting the
$\beta$-function to zero by hand. It is expected to have full symmetries of a conformally
invariant theory. The extra term involving the $\beta$-function can be calculated separately
and e.g. the leading contribution to $\Delta \mathcal{Q}$ can be evaluated very easily
via quark bubble insertions.

A prominent example is provided by the generalized Crewther
relation~\cite{Crewther:1972kn,Broadhurst:1993ru,Crewther:1997ux} between the Bjorken sum rule
in deep inelastic scattering and the total cross section of $e^+e^-$ annihilation:
The expected structure has been confirmed by direct calculations to the $\mathcal{O}(\alpha_s^4)$ accuracy~\cite{Baikov:2010je,Baikov:2012zn}.

Another important application concerns the evolution equation for meson distribution amplitudes and
generalized parton distributions. As shown by D.~M\"uller, off-diagonal terms of the
anomalous dimension matrix of leading twist operators to the $\ell$-loop accuracy are determined by the
special conformal transformation ano\-ma\-ly at one order less~\cite{Mueller:1991gd}. This approach was later used to
calculate the complete two-loop mixing matrix for twist-two operators in
QCD~\cite{Mueller:1993hg,Mueller:1997ak,Belitsky:1997rh},
and derive the two-loop evolution kernels in momentum space for the generalized parton
distributions~\cite{Belitsky:1998vj,Belitsky:1999hf,Belitsky:1998gc}.

In this paper we present an alternative technique to study implications of conformal invariance and illustrate it on two examples:
Calculation of the anomalous dimension matrix in the $O(n)$-symmetric $\varphi^4$ theory in $d=4$ to the three-loop accuracy,
and the $su(n)$ matrix $\varphi^3$ theory in $d=6$ to the two-loop accuracy.
The motivation for our study, apart from various phenomenological applications, is the following.

First of all, we think that the calculations can be considerably simplified by going over to
evolution equations for non-local light-ray operators in position space. In particular the intricate procedure
for the restoration of the evolution kernels from local operators can be avoided.

Second, we want to make the separation of perturbation theory in ``conformal part''
and ``corrections proportional to the $\beta$-function'' as in Eq.~(\ref{conformal}) to be more transparent.
Our starting point is the observation that QCD and toy-model scalar theories that we consider for illustration
possess a nontrivial fixed point in non-integer dimension, $d=4-2\epsilon$
($d=6-2\epsilon$ for $\varphi^3$)~\cite{Banks:1981nn,Hasenfratz:1992jv}.
Conformal symmetry is an \emph{exact} symmetry of the interacting theory for the fine-tuned (critical) value of coupling.
As a consequence, the renormalization group equations are exactly conformally invariant: the
evolution kernels commute with the generators of the conformal group.
The generators are, however, modified by quantum corrections as compared to their canonical
expressions, $S_\alpha=S_\alpha^{(0)}+\Delta S_\alpha$, and the corrections can be calculated order
by order in the perturbative expansion. From the pure technical point of view, this calculation
replaces evaluation of the conformal anomaly in the theory with broken symmetry in integer dimensions
via the Conformal Word identities (CWI) in the approach of D.~M{\"u}ller.
We show that the non-invariant part of the evolution equations with respect to canonical transformations
in $\ell-$th order of perturbation theory is uniquely fixed by the generators in the order $\ell-1$,
and the invariant part is determined (and can be easily restored) from the spectrum of anomalous dimensions.

Last but not least, in $\text{MS}$-like schemes the evolution kernels (anomalous dimensions)
do not depend on the space-time dimension by construction. Thus all expressions derived in the
$d$-dimensional (conformal) theory remain exactly the same for the theory in integer dimensions;
considering the theory at the critical point one does not lose any information.

As already mentioned, the present paper is explo\-ra\-tory.
We work out the necessary formalism for the simplest, scalar field theories.
We expect, however, that the same technique can be used in gauge theories
and in particular in QCD.
The corresponding generalization and applications will be considered elsewhere.

\section{General formalism}
\label{sec:1}
This section is introductory and contains mostly some  general remarks.
\subsection{Scalar field theories}
We will consider the conventional $O(n)$-symmetric $\varphi^4$ theory in $d=4-2\epsilon$ dimensions
\begin{align}\label{Sphi4}
S^{(4)}(\varphi)=\int d^dx\left[\frac12(\partial\varphi)^2+\frac{gM^{2\epsilon}}{24}(\varphi^2)^2\right]\,,
\end{align}
where $\varphi^2=\sum_{a=1}^{n}(\varphi^{a})^2$, and the (somewhat exotic) $su(n)$-matrix
$\varphi^{3}$ theory
\begin{align}\label{Sphi3}
S^{(3)}(\varphi)=&\int d^dx\left[\tr(\partial\varphi)^2+\frac{2}{3}g M^\epsilon \tr\varphi^3\right]\,
\end{align}
in $d=6-2\epsilon$ dimensions. Here $\varphi=\sum_{a}\varphi^a t^a$ and the matrices $t^a$
are the $su(n)$ generators in the fundamental representation normalized as
$\tr t^{a}t^{b}=1/2$. One can  rewrite~(\ref{Sphi3}) as follows
\begin{align}
S^{(3)}(\varphi)
=&\int d^dx\left[\frac12(\partial\varphi^a)^2+\frac{gM^\epsilon}6 d^{abc}\varphi^a\varphi^b\varphi^c\right]\,,
\end{align}
where $d^{abc}=2\tr t^a\{t^b t^c\}$.
Both theories are multiplicatively renormalizable
\begin{align}\label{}
S_R(\varphi)=\int d^d x\left[\frac12Z_1(\partial\varphi)^2+Z_3M^{k\epsilon} g \,V(\varphi)\right]\,,
\end{align}
where $k=1,2$ for  $\varphi^3$ and $\varphi^4$ interaction, respectively.

The renormalization constants $Z_1$ and $Z_3$ for the $\varphi^4$ theory can be found in literature, see e.g.~\cite{AN}:
\begin{align}
Z^{(4)}_1=&1-\frac{u^2}{24\epsilon}\frac{n+2}{3}-\frac{u^3(2-\epsilon)}{48\epsilon^2}\frac{n+2}{3}\frac{n+8}{9}
+\mathcal{O}(u^4),
\notag\\
Z^{(4)}_3
=&1+\frac{u}{\epsilon}\frac{n+8}{6}+\left(\frac{u}{\epsilon}\frac{n+8}{6}\right)^2\!\!-\!
\frac{u^2}{\epsilon}\frac{5n+22}{18}\!+\!\mathcal{O}(u^3).
\end{align}
For the $\varphi^3$ theory we find
\begin{align}
Z^{(3)}_1=&1-\frac{n^2\!-\!4}{2 n}\biggl[\frac{u}{6\epsilon}-\frac{u^2}{36}
\biggl(\frac1{\epsilon^2}\frac{n^2\!-\!16}{ n}
-\frac1\epsilon\frac{n^2\!-\!100}{12 n}\biggr)\biggr]
\notag\\ &
+\mathcal{O}(u^3)\,,
\notag\\
Z^{(3)}_3=&1-\frac{u}{4\epsilon}\frac{n^2-12}{n}+\frac{u^2}{16}
\biggl(\frac1{\epsilon^2}\frac{n^2-4}{n}\frac{n^2-12}{n}
\notag\\
&-\frac1\epsilon
\frac{n^4-100n^2+960}{6n^2}\biggr)+\mathcal{O}(u^3)\,,
\end{align}
where
\begin{align}
  u =\frac{g}{(4\pi)^2} \ [\varphi^4\,\text{theory}],
\qquad
 u = \frac{g^2}{(4\pi)^3} \ [\varphi^3\,\text{theory}].
\end{align}
The beta-function and the anomalous dimension of the basic field are defined as follows
\begin{align}
\beta(u)=\frac{du}{d \ln M}\,,\qquad
\gamma_\varphi=\frac12 \frac{d\ln Z_1}{d\ln M }\,.
\end{align}
One obtains
\begin{align}\label{rgf-4}
\beta(u)=&-2\epsilon u+\frac{u^2(n+8)}3-\frac{u^3(3n+14)}{3}+\mathcal{O}(u^4)\,,
\notag\\
\gamma_\varphi=&\frac{u^2(n+2)}{36}\left[1-\frac{u(n+8)}{12}+\mathcal{O}(u^2)\right]\,
\end{align}
and
\begin{align}
\beta(u)=&-2\epsilon u-u^2\frac{n^2-20}{2n}
\notag\\&-u^3\frac{5n^4-496n^2+5360}{72 n^2}+\mathcal{O}(u^4)\,,
\notag\\
\gamma_\varphi=&u\frac{n^2-4}{12 n}\left(1+u\frac{n^2-100}{36 n}\right)+\mathcal{O}(u^3)\,
\label{rgf-3}
\end{align}
for the $\varphi^4$ and $\varphi^3$ theories, respectively.

The critical coupling is defined by the condition $\beta(u_*) =0$. Solving this equation
for $u_*$, one obtains the well-known expansion for the critical coupling $u_*$ in
powers of $4-d$ in the $\varphi^4$ theory
\begin{align}\label{ust4}
u^{(\varphi^4)}_*=&\frac{6\epsilon}{n+8}+\left(\frac{6\epsilon}{n+8}\right)^2
\frac{3n+14}{n+8}+\mathcal{O}(\epsilon^3)\,,
\end{align}
whereas for the $\varphi^3$ theory we obtain
\begin{align}\label{ust3}
u^{(\varphi^3)}_*\!=\frac{4n\epsilon}{20-n^2}+\frac{4n\epsilon^2}{9}\frac{(5n^4-496n^2+5360)}{(20-n^2)^3}+O(\epsilon^3).
\end{align}
In the latter case a nontrivial critical point for $d<6$ only exists for $n=3$ and $n=4$.
(For $n=2$ the theory is free as the $d_{abc}$-symbols vanish identically). Staying within
perturbation theory on can, however, consider $n$ as a continuous parameter. In this sense
all further results  hold for arbitrary $n$.

Renormalization ensures finiteness of the correlation functions of the basic field that are encoded
in the partition function
\begin{align}
Z(A)=\mathcal{N}^{-1}\int D\varphi \, e^{-S_R(\varphi)+\int d^d x A(x)\varphi(x)}\,.
\label{partition1}
\end{align}
Here $A$ is an external source and, as usual, the normalization is chosen in such a way
that $Z(0)=1$. Correlation functions with an insertion of a composite operator,
$\mathcal{O}_k$, 
possess additional divergences that are removed by the operator renormalization,
\begin{align}
{}[\mathcal{O}_k]=\sum_j Z_{kj} \mathcal{O}_j,
\end{align}
where the sum goes over all operators with the same quantum numbers that get mixed.
Here and below we use square brackets to denote renormalized composite operators (in a minimal subtraction scheme).
%

\subsection{Light-ray operators}\label{sec:nonlocal}
Light-ray operators~(see e.g.~\cite{Balitsky:1987bk})
will always be understood here as generating functions for the leading-twist local operators:
\begin{eqnarray}
 \mathcal{O}^{ab}(x;z_1,z_2) &\equiv& \varphi^a(x+z_1n)\varphi^b(x+z_2n)
\nonumber\\&\equiv& \sum_{mk} z_1^m z_2^k \mathcal{O}_{mk}^{ab}(x)\,,
\label{LRO}
\end{eqnarray}
where
\begin{eqnarray}
 \mathcal{O}_{mk}^{ab}(x)=\frac{1}{m!k!}(n\partial)^m\varphi^a(x)(n\partial)^k\varphi^b(x)\,.
\end{eqnarray}
Here $n^\mu$ is an auxiliary light-like vector, $n^2=0$, that ensures
symmetrization and subtraction of traces of local operators.

A renormalized light-ray operator is the generating
function for renormalized local operators. It can be written in the form
\begin{multline}
{}[\mathcal{O}^{ab}(x;z_1,z_2)] =\sum_{mk} z_1^m z_2^k [\mathcal{O}_{mk}^{ab}(x)]=
\\
=Z^{ab}_{a'b'}\,\mathcal{O}^{a'b'}(x;z_1,z_2)\,,
\end{multline}
where $Z$ (the renormalization constant) is an integral operator
acting on the coordinates $z_1,z_2$ that has an expansion
in inverse powers of $\epsilon$
\begin{align}
Z=1+\sum_{k=1}^\infty \epsilon^{-k} Z_k(u)\,.
\label{Zk}
\end{align}
It is a matrix in isotopic space.
The renormalized light-ray
operator $[\mathcal{O}(x;z_1,z_2)]$ satisfies the renormalization-group (RG) equation
\begin{align}\label{RGO}
\Big(M{\partial_M}+\beta(u)\partial_u +\mathbb{H}\Big)[\mathcal{O}(x;z_1,z_2)]=0\,,
\end{align}
where we suppressed the isotopic indices.
The coupling $u$ for the theories in question is defined in Eq.(\ref{rgf-3})
and the evolution kernel (Hamiltonian) $\mathbb{H}$ is given by
\begin{align}
\mathbb{H}=-\left({M}\frac{d}{dM} \mathbb{Z}\right) \mathbb{Z}^{-1}=2u\partial_u Z_1(u)+2\gamma_\varphi\,,
\label{H}
\end{align}
where $\mathbb{Z}=Z Z_1^{-1}$. In perturbation theory $\mathbb{H}$ can be written as
a series
\begin{align}\label{Hexp}
\mathbb{H}=u\mathbb{H}^{(1)}+u^2\mathbb{H}^{(2)}+\ldots\,.
\end{align}
The kernels $\mathbb{H}^{(k)}$ in minimal subtraction schemes
do not depend $\epsilon$. As a consequence these kernels are
exactly the same for the theories in $d$ dimensions that we consider at the
intermediate step and physical theories in integer dimensions that  are our final goal.
We stress that Eq.~(\ref{RGO}) is completely equivalent to the RG equation for the
local twist-two operators,
\begin{align}
\Big([M{\partial_M}+\beta(u)\partial_u]\delta_{mk}^{m'k'}\delta^{ab}_{a'b'} +
(\gamma_{mk}^{m'k'})^{ab}_{a'b'}\Big)[\mathcal{O}_{m'k'}^{a'b'}]=0\,,
\end{align}
where $\gamma$ is the usual anomalous dimension matrix.

\subsection{Conformal symmetry}

The usual Poincare symmetry of the theory is enhanced at the critical point  $u=u_*$,
$\beta(u_*)=0$ by the dilatation (scale invariance) and space-time inversion.
For our purposes it is sufficient to consider the transformations that act nontrivially
on the twist-two (symmetric and traceless) operators. These transformations
form the so-called collinear $sl(2)$ subgroup of the full conformal group that
leaves the light-ray $x^\mu=z n^\mu $ invariant, see Ref.~\cite{Braun:2003rp} for a review.

Collinear conformal transformations are generated by translations along the light-ray
direction $n^\mu$, special conformal transformations in the alternative light-like direction $\bar n$,
$\bar n^2=0$, $(n\bar n)=1$, and the combination of the dilation and rotation in the $(n,\bar n)$ plane
\begin{align}
\mathbf{L}_-=-i\mathbf{P}_n,\! &&\!\mathbf{L}_+=\frac12i\mathbf{K}_{\bar n},\! &&\!
\mathbf{L_0}=\frac{i}2\left(\mathbf{D}-\mathbf{M}_{n\bar n}\right).
\label{gene}
\end{align}
Here and below we use a shorthand notation $\mathbf{P}_n = n^\mu \mathbf{P}_\mu$ etc.
The generators defined in this way satisfy standard $sl(2)$ commutation
relations
\begin{align}\label{}
{}[\mathbf{L}_{\pm},\mathbf{L}_0]=\pm \mathbf{L}_{\pm}\,, &&
{}[\mathbf{L}_{+},\mathbf{L}_-]=-2\mathbf{L}_0\,.
\end{align}

Local composite operators can be classified according to irreducible representations of the $sl(2)$
algebra. A traceless and symmetric (renormalized) operator
$$
[\mathcal{O}_N](x)=n_{\mu_1}\ldots n_{\mu_N} [\mathcal{O}^{\mu_1\ldots\mu_N}_N](x),
$$
(for a while we suppress isotopic indices) is called conformal if it transforms covariantly under
the special conformal transformation~\cite{Derkachov:1992he}:
\begin{multline}
i\big[\mathbf{K}^\mu,[\mathcal{O}_N](x)\big]=
\biggl[2 x^\mu (x
\partial)-x^2\partial^\mu+2\Delta^\ast_N x^\mu
\\
+2x^\nu \left(n^\mu \frac{\partial}{\partial n^\nu}- n_\nu \frac{\partial}{\partial n_\mu}
\right)\biggr][\mathcal{O}_N](x)\,.
\end{multline}
Here $\Delta^\ast_N$ is the scaling dimension of the operator (at the critical point):
\begin{align}
i\big[\mathbf{D},[\mathcal{O}_N](x)\big]=\big(x\partial_x+\Delta^\ast_N\big)[\mathcal{O}_N](x)\,.
\end{align}
As a consequence of having definite scaling dimension, the conformal operator $\mathcal{O}_{N}$
satisfies the RG equation %
\begin{align}\label{OM}
\Big(M{\partial_M}+\gamma^*_N\Big)[\mathcal{O}_{N}]=0\,,
\end{align}
where $\gamma_N^*$ is the anomalous dimension at the critical point,
$\gamma_N^*=\gamma_N(u_*)$. The scaling dimension is given by the sum of the canonical and
anomalous dimensions, $\Delta^\ast_N=\Delta_N+\gamma_N^*$. For the operators under
consideration $\Delta_N=2\Delta+N$ where $\Delta=d/2-1$ is the canonical dimension
of the basic field $\varphi(x)$.

Each conformal operator \ $[\mathcal{O}_N]$
generates an irreducible representation of the $sl(2)$ algebra (conformal tower),
consisting of operators obtained by adding total derivatives:
\begin{align}
 \mathcal{O}_{Nk}=(n\partial)^k [\mathcal{O}_N(0)],\,\qquad k=0,1,\ldots
\end{align}
such that
\begin{align}\label{3L}
\delta_-\mathcal{O}_{Nk}=~&\big[\mathbf{L}_-,\mathcal{O}_{Nk}\big]=~-\mathcal{O}_{Nk+1}\,,
\notag\\
\delta_{0\phantom{i}}\mathcal{O}_{Nk}=~&\big[\mathbf{L}_0,\,\mathcal{O}_{Nk}\big]=~(j_N+k)\mathcal{O}_{Nk}\,,
\notag\\
\delta_+\mathcal{O}_{Nk}=~&\big[\mathbf{L}_+,\mathcal{O}_{Nk}\big]=~k(2j_N+k-1)\mathcal{O}_{Nk-1}\,,
\end{align}
with the operator $[\mathcal{O}_N]$ itself being the highest weight vector,
$\big[\mathbf{L}_+,[\mathcal{O}_{N}]\big]=0$. Here $j_N$ is the so-called conformal spin of
the operator --- the half-sum of its scaling dimension and spin
\begin{align}
j_N=\frac12(\Delta^\ast_N+N) = \Delta + N  + \frac12 \gamma_N^*.
\end{align}
All operators $\mathcal{O}_{Nk}$ in a conformal tower have, obviously, the same anomalous
dimension $\gamma_N^\ast$.

Going over from the description in terms of conformal towers of local operators to the
light-ray operators essentially corresponds to going over to a different realization of conformal symmetry.
Due to Poincare invariance one can put, without loss of generality, $x=0$ in a definition on light-ray
operator~(\ref{LRO}). Hereafter we consider $$[\mathcal{O}(z_1,z_2)]\equiv [\mathcal{O}(x=0;z_1,z_2)].$$
The light-ray operator $[\mathcal{O}(z_1,z_2)]$ can be expanded in terms of local operators~$\mathcal{O}_{Nk}$
\begin{align}\label{nlo2}
[\mathcal{O}(z_1,z_2)]=\sum_{Nk}\Psi_{Nk}(z_1,z_2)\,[\mathcal{O}_{Nk}]\,,
\end{align}
where $\Psi_{Nk}(z_1,z_2)$ are homogeneous polynomials of degree $N+k$
\begin{align}
  \left(z_1\partial_{z_1} + z_2\partial_{z_2} -N-k\right)\Psi_{Nk}(z_1,z_2) =0\,.
\end{align}
The action of the generators $\mathbf{L}_{\pm,0}$ on the quantum fields in the light-ray operator
can be traded for the differential operators $S_{\pm,0}$ acting on the field coordinates, i.e. on the coefficient functions $\Psi_{Nk}(z_1,z_2)$:
\begin{align}
 \delta_{\pm,0} [\mathcal{O}(z_1,z_2)] =  S_{\pm,0} [\mathcal{O}(z_1,z_2)]\,.
\end{align}
The generators $S_{\pm,0}$ obey the usual $sl(2)$ commutation relations
\begin{align}\label{sl2-comm}
{}[S_0,S_{\pm}]=\pm S_{\pm}\,, &&
{}[S_{+},S_-]= 2S_0\,
\end{align}
and their action on the coefficient functions in the expansion (\ref{nlo2}) takes the form
\begin{align}
   S_- \Psi_{Nk}(z_1,z_2) =& - \Psi_{Nk-1}(z_1,z_2)\,,
\notag\\
   S_0\, \Psi_{Nk}(z_1,z_2) =& (j_N+k) \Psi_{Nk}(z_1,z_2)\,,
\notag\\
   S_+ \Psi_{Nk}(z_1,z_2) =&  (k+1)(2j_N+k)\Psi_{Nk+1}(z_1,z_2)\,.
\label{steps}
\end{align}
With the exception of $S_-$, the form of the generators in the interacting theory (at the critical point) differs, however, from the
canonical expressions (see e.g. \cite{Braun:2003rp})
\begin{align}\label{S_0}
S_-^{(0)}=&-\partial_{z_1}-\partial_{z_2}\,,
\notag\\
S_0^{(0)}=&~z_1\partial_{z_1}+z_2\partial_{z_2}+2j\,,
\notag\\
S_+^{(0)}=&~z_1^2\partial_{z_1}+z_2^2\partial_{z_2}+2j(z_1+z_2)\,.
\end{align}
Here $j$ is the conformal spin of the field $\varphi$, $j=1/2$ ($\varphi^4$-theory) and
$j=1$ ($\varphi^3$-theory).
We obtain (see below)
\begin{align}
  S_- =& S_-^{(0)}\,,
\notag\\
  S_0\, =& S_0^{(0)} + \Delta S_0^{(0)}\,,
\notag\\
  S_+ =& S_+^{(0)} + \Delta S_+^{(0)}\,,
\label{DeltaS}
\end{align}
where
\begin{align}\label{DeltaS0+}
\Delta S_0=& -\epsilon+\frac12 \mathbb{H}(u_*)\,.
\notag\\
\Delta S_+=&(z_1+z_2)\Big(-\epsilon+\frac12 u_*\, \mathbb{H}^{(1)}\Big)+\mathcal{O}(\epsilon^2)\,.
\end{align}
Note that $\Delta S_0$ is given in closed form in terms of the evolution kernel $\mathbb{H}$,
whereas $\Delta S_+$ can only be calculated as a series expansion in $\epsilon$ and/or $u_\ast =\mathcal{O}(\epsilon)$.
To the $\mathcal{O}(\epsilon)$ accuracy the result (given above) turns out to be the same in both
theories that we consider in this work. For the case of the $\varphi^4$ theory we have also calculated
the next, $\mathcal{O}(\epsilon^2)$, correction.

If $\Delta S_+$ is known (to a given order), one can use the last equation in (\ref{steps})
to construct the whole set of the coefficient functions $\Psi_{Nk}$ to the same accuracy starting from
the lowest one, $\Psi_{Nk=0}=(z_1-z_2)^N$. This in turn is sufficient in order to be able
to obtain explicit expressions for the multiplicatively renormalizable operators
$\mathcal{O}_{Nk}$ in terms of the operators $\mathcal{O}_{nm}$,
cf. Eq.~(\ref{LRO}), by comparing the coefficients of $z_1^n z_2^m$.
Thus the operator $S_+$ effectively encodes all information on the form of the eigenoperators
of the evolution equation at the critical point.

Conformal symmetry implies that the full evolution kernel for light-ray operators for the critical coupling
\begin{align}\label{Hcrit}
\mathbb{H}(u_\ast) =u_\ast\,\mathbb{H}^{(1)}+u_\ast^2\,\mathbb{H}^{(2)}+\ldots\,
\end{align}
commutes with the exact $sl(2)$ generators
\begin{align}
{}[\mathbb{H}(u_\ast),S_{\pm,0}] = 0\,.
\label{com22}
\end{align}
Hence the evolution kernels at each order in perturbation theory, $\mathbb{H}^{(k)}$, can be
split in the $sl(2)$-invariant and non-invariant parts with respect to the canonical transformations (\ref{S_0})
\begin{align}
  \mathbb{H}^{(k)} = \mathbb{H}^{(k)}_{\rm inv}+\Delta \mathbb{H}^{(k)}\,,
\label{inv-noninv}
\end{align}
such that
\begin{align}
{}[\mathbb{H}^{(k)}_{\rm inv},S^{(0)}_{\pm,0}] = 0\,.
\end{align}
We will show that the non-invariant part of the $k$-loop kernels
$\Delta \mathbb{H}^{(k)}$ is uniquely determined by the $(k-1)$-loop result for $S_+$,
after which the invariant part $\mathbb{H}^{(k)}_{\rm inv}$ can easily be restored from the anomalous dimensions.

\section{Deformed $sl(2)$ generators}

Thanks to Poincare invariance the  generator $S_-$ does not receive any corrections
in the interacting theory, i.e. $S_-=S_-^{(0)}$.

Indeed, since $\mathcal{O}(x;z_1,z_2)$ actually depends on the two field coordinates
$x+z_1n$ and $x+z_2n$, a translation along the light ray $x^\mu \to x^\mu + a n^\mu$
can be compensated by redefinition of the $z$-coordinates $z_{1,2} \to z_{1,2} -a$.
This means that action of the quantum operator $\mathbf{L}_-=-i\mathbf{P}_n$ on
the quantum fields in $\mathcal{O}(z_1,z_2)$ can be traded for the shift in the field coordinates
\begin{eqnarray}
\delta_- [\mathcal{O}(z_1,z_2)] &\equiv& \big[\mathbf{L}_-, [\mathcal{O}(z_1,z_2)]\big]
\nonumber\\ &=& -(\partial_{z_1}+\partial_{z_2})[\mathcal{O}(z_1,z_2)]\,.
\end{eqnarray}
Since, on the other hand
\begin{eqnarray}
\delta_- [\mathcal{O}(z_1,z_2)] &=& - \sum_{Nk}\Psi_{Nk}(z_1,z_2)\,[\mathcal{O}_{N,k+1}]\,,
\label{dminus2}
\end{eqnarray}
we conclude that $S_- = - (\partial_{z_1}+\partial_{z_2})$ acts as a step-down operator
in the space of coefficient functions, cf. the first Eq.~(\ref{steps}).
Note that the expansion on the r.h.s. in (\ref{dminus2}) starts with $\mathcal{O}_{Nk=1}$ which means that
the coefficient function of the conformal operator  $\mathcal{O}_{N} \equiv \mathcal{O}_{Nk=0}$ is annihilated
by $S_-$. Hence
\begin{eqnarray}
  \Psi_{N}(z_1,z_2) \equiv \Psi_{N0}(z_1,z_2) = c_N (z_1-z_2)^N \equiv c_N z_{12}^N \,,
\end{eqnarray}
where the coefficients $c_N$ depend on the normalization convention for the conformal operators.

Next, let us consider $S_0$. Using Eq.~(\ref{3L}) we obtain
\begin{eqnarray}
\delta_0 [\mathcal{O}(z_1,z_2)] &\equiv& \big[\mathbf{L}_0, [\mathcal{O}(z_1,z_2)]\big]
\nonumber\\ &=&
\sum_{Nk}\Psi_{Nk}(z_1,z_2)\,(j_N+k)[\mathcal{O}_{N,k}]\,,
\end{eqnarray}
where one can rewrite
\begin{align}
j_N+k=N+k+\Delta+\frac12\gamma_N^*\,.
\end{align}
It follows from Eqs.~(\ref{RGO}) and (\ref{OM}) that the functions $\Psi_{Nk}(z_1,z_2)$
are the eigenfunctions of the evolution kernel $\mathbb{H}$ for the critical value of
coupling
\begin{align}\label{EigenPsi}
[\mathbb{H}(u_*)\Psi_{Nk}](z_1,z_2)=\gamma_N^*\,\Psi_{Nk}(z_1,z_2)\,.
\end{align}
Thus one obtains
\begin{eqnarray}
  \delta_0 [\mathcal{O}(z_1,z_2)] & = & S_0 [\mathcal{O}(z_1,z_2)]
\end{eqnarray}
with
\begin{eqnarray}\label{S0}
  S_0 &=& z_1\partial_{z_1}+z_2\partial_{z_2} +\Delta+\frac12\mathbb{H}(u_*) \,,
\end{eqnarray}
which is the result quoted in Eq.~(\ref{DeltaS0+}).

Unfortunately, the deformation of the generator $S_+$ in interacting theory cannot be found using
similar general arguments and requires an explicit calculation. It can be done using the special conformal
Ward identity. To this end we consider the partition function with
the insertion of the renormalized light-ray operator $[\mathcal{O}(z_1,z_2)]$, cf. Eq.~(\ref{partition1}):
\begin{eqnarray}\label{Gc}
Z_{\mathcal{O}}(z_1,z_2;A)&=&\VVV{[\mathcal{O}(z_1,z_2)]}=
\nonumber\\&=&
\mathcal{N}^{-1}\!\int D\varphi \,[\mathcal{O}(z_1,z_2)]
e^{-S_R(\varphi)+A\varphi}\,.
\end{eqnarray}
Let us make a change of variables in the functional integral~(\ref{Gc}):
\begin{align}\label{ftf}
\varphi(x)\to
\varphi(x)+\delta_c\varphi(x)=\varphi(x)+\omega K_{\bar n}(\Delta)\varphi(x)\,.
\end{align}
Here $\omega$ is a small parameter and the  operator $K_{\bar n} = \bar n^\mu K_\mu$ is
the generator  of special conformal transformations
\begin{align}
K_\mu(\Delta)=2x_\mu(x\partial)-x^2\partial_\mu+2\Delta x_\mu\,.
\label{Kmu}
\end{align}
Note that we have chosen  the parameter $\Delta$ entering the definition of $K_\mu$  equal
to the canonical dimension of the field~$\varphi$, $\Delta=d/2-1$. With this choice the
kinetic term in the action is invariant under special conformal transformations.

Further, note  that $\delta_c[\mathcal{O}(z_1,z_2)]=Z\delta_c\mathcal{O}(z_1,z_2)$.
Variation of the bare light-ray operator can easily be calculated using the
definition in Eq.~(\ref{Kmu}):
\begin{align}
\delta_c\mathcal{O}(z_1,z_2)=2\omega(n\bar n) \bar S_+ \mathcal{O}(z_1,z_2)\,,
\end{align}
where
\begin{align}
\bar S_+=z_1^2\partial_{z_1}+z_{2}^2\partial_{z_2}+\Delta(z_1+z_2) = S^{(0)}_+ -\epsilon(z_1+z_2)\,.
\end{align}
Thus
\begin{align}
\delta_c[\mathcal{O}(z_1,z_2)]=2\omega(n\bar n) Z\bar S_+ Z^{-1}[\mathcal{O}(z_1,z_2)]\,.
\end{align}
Since the partition function (\ref{Gc}) does not change under the change of variables (\ref{ftf}),
one obtains an identity
\begin{eqnarray}\label{CWI0}
0&=& \Big(\int\! d^dy\, A(y) K^{y}_{\bar n}(\Delta)\frac{\delta}{\delta A(y)}
+2(n \bar n)Z\bar S_+ Z^{-1}\Big)
\nonumber\\&&{}
\times Z_{\mathcal{O}}(z_1,z_2;A)
-\VVV{\delta_c S_R(\varphi)[\mathcal{O}(z_1,z_2)]}\,,
\end{eqnarray}
where the superscript in $K^{y}_{\bar n}$  indicates
the variable the operator acts on
and
$\delta_c S_R(\varphi)$ is the variation of the action
under the special conformal transformation~(\ref{ftf})
\begin{align}\label{deltaSE}
\delta_c S_R(\varphi) =&-\frac13 \epsilon
{Z_3 M^\epsilon g}
\int\! d^d x\, x_{\bar n}d^{abc}\varphi^a(x)\varphi^b(x)\varphi^c(x)\,,
\notag\\
\delta_c S_R(\varphi) =&-\frac16 \epsilon
{Z_3M^{2\epsilon} g}
\int\! d^d x\, x_{\bar n}\varphi^4(x)
\end{align}
for the $\varphi^3$- and $\varphi^4$-theories, respectively.

It should be stressed that the CWI~(\ref{CWI0}) holds for arbitrary value of the coupling
constant. Note also that since it is derived by a variation of the finite (renormalized)
partition function, all singular $1/\epsilon^k$-terms  in Eq.~(\ref{CWI0}) have to cancel each other.

Conformal symmetry of the theory at the critical point implies that Eq.~(\ref{CWI0}) can be rewritten in the form
\begin{eqnarray}\label{CWIC}
0 &=& \Big(\int d^dy\, A(y) K^{y}_{\bar n}(\Delta^\ast)\frac{\delta}{\delta A(y)}+2(n \bar n) S_+ \Big)
\nonumber\\&&{}\times
Z_{\mathcal{O}}(z_1,z_2;A)\,,
\end{eqnarray}
where $\Delta^*$ is the critical scaling dimension of the field $\varphi$, $\Delta^* = \Delta +\gamma^\ast_\varphi$.
Evaluating the entries in Eq.~(\ref{CWI0}) in perturbation theory and bringing the result to the form (\ref{CWIC}) one obtains
the generator $S_+$ as a series in $\epsilon$ or, equivalently,  $u_\ast(\epsilon)$.

\begin{figure}
\centerline{ \includegraphics[width=0.320\textwidth]{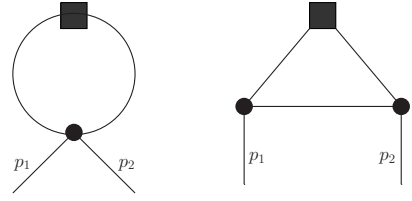}}
\caption{One-loop diagrams for the 1PI Green function $\Gamma_2(\underline{z},\underline{p})$
in the $\varphi^4$-theory (left) and $\varphi^3$-theory (right).
The boxes denote the insertion of the light-ray operator $[\mathcal{O}(z_1,z_2)]$.}
\label{fig:1}
\end{figure}

It turns out to be more convenient to analyze the corresponding identities for one-particle irreducible (1PI)
Green functions, $\Gamma_{\mathcal{O}}(z_1,z_2,\varphi)$.
The CWIs~(\ref{CWI0}) and (\ref{CWIC}) are replaced in this case by
\begin{eqnarray}\label{CWI0a}
0&=& \Big(\int\! d^dy\, \varphi(y) K^{y}_{\bar n}(\widetilde\Delta)\frac{\delta}{\delta \varphi(y)}
+2(n \bar n)Z\bar S_+ Z^{-1}\Big)
\nonumber\\&&{}
\times \Gamma_{\mathcal{O}}(z_1,z_2;\varphi)
-\VVV{\delta_c S_R(\varphi)[\mathcal{O}(z_1,z_2)]}_{1PI}\,,
\end{eqnarray}
\begin{eqnarray}\label{CWICa}
0 &=& \Big(\int d^dy\, \varphi(y) K^{y}_{\bar n}(\widetilde\Delta^\ast)\frac{\delta}{\delta \varphi(y)}+2(n \bar n) S_+ \Big)
\nonumber\\&&{}\times
\Gamma_{\mathcal{O}}(z_1,z_2;\varphi)\,,
\end{eqnarray}
respectively, where the $\widetilde \Delta$ is the shadow dimension $$\widetilde
\Delta=d-\Delta.$$

As the first step, let us rewrite
\begin{align}\label{Su}
S_+(u)\equiv Z(u)\bar S_+ Z^{-1}(u)=\mathbb{Z}(u)\bar S_+ \mathbb{Z}^{-1}(u)\,,
\end{align}
where $\mathbb{Z}=Z Z_1^{-1}$ in terms of the evolution kernel $\mathbb{H}$ (\ref{H}).
To this end, note that $S_+(u)$ obeys the following differential equation:
\begin{align}
M\frac{d}{dM}S_+(u)=\beta(u)\frac{d}{du}S_+(u)=-[\mathbb{H}(u),\,S_+(u)]\,,
\end{align}
which follows readily from Eq.~(\ref{Su}). Taking into account that in the free theory
$S_+(u=0)=\bar S_+$ one obtains
\begin{multline}\label{Sexp}
S_+(u)=\bar S_+-\int_0^{u}\frac{dv}{\beta(v)}[\mathbb{H}(v),\bar S_+]\\
+ \int_0^{u}\frac{dv}{\beta(v)}\int_0^v
\frac{dw}{\beta(w)}[\mathbb{H}(v),[\mathbb{H}(w),\bar S_+]]+ \ldots
\end{multline}
Substituting in this equation the evolution kernel $\mathbb{H}(u)$
and the beta-function $\beta(u)$ by their perturbative expansion
\begin{align}
\mathbb{H}(u)=&~u\,\mathbb{H}^{(1)}+u^2\,\mathbb{H}^{(2)}+\ldots
\notag\\
\beta(u)=&u\big(-2\epsilon  + u\beta_0+u^2\beta_1+\ldots\big)
\end{align}
we get
\begin{eqnarray}\label{Su+}
S_+(u)&=&\bar S_++\frac{u}{2\epsilon}[\mathbb{H}^{(1)},\bar S_+]\left(1+\frac{\beta_0
u}{4\epsilon}\right) +\frac{u^2}{4\epsilon}[\mathbb{H}^{(2)},\bar S_+]
\nonumber\\&&{}
+\frac{u^2}{8\epsilon^2}[\mathbb{H}^{(1)}[\mathbb{H}^{(1)},\bar S_+]]+\mathcal{O}(u^3)
\end{eqnarray}
Taking into account that the leading-order evolution kernel $\mathbb{H}^{(1)}$ commutes with the
canonical generator $S_+^{(0)}$ (\ref{S_0}) and writing $\bar S_+ = S_+^{(0)}-\epsilon(z_1+z_2)$
one can further simplify Eq.~(\ref{Su+}) as follows:
\begin{align}\label{S+exp}
S_+(u)=& \bar S_+-\frac{u}{2}[\mathbb{H}^{(1)},(z_1+z_2)]
-\frac{u^2}{4}[\mathbb{H}^{(2)},(z_1+z_2)]
\notag\\
&
-\frac{u^2}{8\epsilon}
\biggl\{\beta_0[\mathbb{H}^{(1)},(z_1+z_2)]+2[S_+^{(0)},\mathbb{H}^{(2)}]
\notag\\
&
+[\mathbb{H}^{(1)}[\mathbb{H}^{(1)},(z_1+z_2)]]
\biggr\}+\mathcal{O}(u^3)\,.
\end{align}
Note that this expression contains both regular and singular parts in $1/\epsilon$,
$S_+(u)=S^{(\text{reg})}_+(u)+S^{(\text{sing})}_+(u)$.
It is easy to see that the regular part comes solely from the first two terms in the expansion~(\ref{Sexp})
so that to all orders in the coupling
\begin{align}
S^{(\text{reg})}_+(u)=S_+^{(0)}-\epsilon(z_1\!+\!z_2)-\frac12\int_{0}^u\frac{dv}v[\mathbb{H}(v),z_1\!+\!z_2]\,.
\end{align}
The singular part, $S^{(\text{sing})}_+(u)$, receives contributions from all terms in the expansion~(\ref{Sexp}).
Since all $\epsilon$-singular terms in the CWI~(\ref{CWI0a}) must cancel,
the sum of the last two terms in this identity, $2(n\bar n) S_+(u)\Gamma_\mathcal{O}(z_1,z_2,\varphi)$ and
$\VVV{\delta_c S_R(\varphi)[\mathcal{O}(z_1,z_2)]}_{1PI}$, has to be finite.
This implies that the counterterm to the operator $\delta_c S_R(\varphi)[\mathcal{O}(z_1,z_2)]$
must have the following form:
\begin{align}
2(n\bar n) S^{(\text{sing})}_+(u)[\mathcal{O}(z_1,z_2)]\,.
\end{align}
As seen from  Eq.~(\ref{S+exp}) $S^{(\text{sing})}_+(u)\sim \mathcal{O}(u^2)$ and
thus the correlation function $\VVV{\delta_c S_R(\varphi)[\mathcal{O}(z_1,z_2)]}_{1PI}$ must be finite
to the leading order in $u$.

Let us analyze this contribution in detail. To this end it is sufficient to consider the
two-point $1PI$ function
\begin{multline}\label{DeltaSC}
\VEV{\delta_cS_R(\varphi)\,[\mathcal{O}(z_1,z_2)]\varphi(x_1)\varphi(x_2)}_{1PI}=
\\=\int d^dp_1 d^d p_2 \, e^{-ip_1x_1-ip_2 x_2}
\delta\Gamma_{2}(z_1,z_2,p_1,p_2)\,.
\end{multline}
Let also
\begin{multline}
\Gamma_{2}(z_1,z_2,x_1,x_2) = \VEV{[\mathcal{O}(z_1,z_2)]\varphi(x_1)\varphi(x_2)}_{1PI}
\\=\int d^dp_1 d^d p_2 e^{-ip_1x_1-ip_2 x_2}
\Gamma_{2}(z_1,z_2,p_1,p_2)\,.
\label{Gamma2}
\end{multline}
To save space below we use a shorthand notation for the arguments
$$\underline{z} = \{z_1,z_2\}\,\qquad \underline{p} = \{p_1,p_2\}\,, $$
etc.

As follows from the explicit expressions in Eq.~(\ref{deltaSE}), variation of the action can be
written in both theories as
\begin{align}\label{Veff}
\delta_c S_R(\varphi)=-2k\epsilon Z_3 M^{k\epsilon} g\int d^dx\, (\bar n x) V(\varphi),
\end{align}
where $k=1,2$ for $\varphi^3-$ and $\varphi^4$-theories, respectively, and
$V(\varphi)$ is the corresponding potential.

The one-loop Feynman diagrams for $\delta
\Gamma_2(\underline{z},\underline{p})$ are obtained from the diagrams shown in Fig.~\ref{fig:1}
by the replacement of one of the interaction vertices by an effective
vertex derived from~(\ref{Veff}). The factor $(\bar n x)$ in the effective vertex can be
represented as the derivative with respect to the incoming momentum~\cite{Mueller:1991gd}
$$
\int d^dx\, e^{i(px)}(\bar n x)V(\varphi)=-i(\bar n \partial_p)\int d^dx\,
e^{i(px)}V(\varphi)\,.
$$
Thus to the one-loop accuracy
\begin{align*}
\delta\Gamma^{(1)}_{2}(\underline{z},\underline{p})=
-2i\epsilon\Big(\bar n,\frac{\partial}{\partial p_1}+\frac{\partial}{\partial p_2}\Big)
\Gamma^{(1,{\rm bare})}_{2}(\underline{z},\underline{p})\,,
\end{align*}
where $(*,*)$ is the usual Minkowski scalar product and
$\Gamma^{(1,{\rm bare})}_{2}$ is the unrenormalized Green function,
$\Gamma^{(1)}_{2}=\Gamma^{(1,{\rm bare})}_{2}-\,\text{counterterm}$.
Note that this relation holds for the both theories considered here. Taking into account that
$\Gamma^{(1,{\rm bare})}_{2}\sim M^{2\epsilon}$ and that the tree-level function
$\Gamma^{(0)}_2(\underline{z},\underline{p})$ (and, hence, the counterterm)
does not depend on $M$, we can rewrite this expression as
\begin{eqnarray}
\lefteqn{\delta\Gamma^{(1)}_{2}(\underline{z},\underline{p})=}
\nonumber\\&=&
-i (\bar n,\partial_{p_1}+\partial_{p_2})M\partial_M\Gamma^{(1)}_2(\underline{z},\underline{p})
\nonumber\\&=&
i(\bar n,\partial_{p_1}+\partial_{p_2})\big[\beta(u)\partial_u + \mathbb{H}-2\gamma_\varphi\big]
\Gamma^{(0)}_2(\underline{z},\underline{p}).
\end{eqnarray}
Since $\Gamma^{(0)}_2(\underline{z},\underline{p})\sim e^{i(n,p_1 z_1+p_2 z_2)}$
the derivatives in the momenta are trivial so that
\begin{eqnarray}
\lefteqn{\delta\Gamma^{(1\ell)}_{2}(\underline{z},\underline{p})=}
\nonumber\\&=&-(n\bar n)\big[\beta(u)\partial_u+ \mathbb{H}-2\gamma_\varphi\big]
(z_1+z_2) \Gamma^{(0)}_2(\underline{z},\underline{p})\,.
\end{eqnarray}
The tree-level 1PI function in coordinate space is given by the product of
delta-functions, $$\Gamma^{(0)}_2(\underline{z},\underline{x})\sim
\delta(x_1-z_1n)\delta(x_2-z_2 n),$$
hence
\begin{eqnarray}
\label{DScorr}
\lefteqn{\VEV{\delta_cS_R(\varphi)\,[\mathcal{O}(z_1,z_2)]\varphi(x_1)\varphi(x_2)}_{1PI}=}
\nonumber\\&=&
\Big\{\gamma_\varphi(x_1 +x_2,\bar n)
-(n\bar n)\big[\beta(u)\partial_u+\mathbb{H}\,(z_1+z_2)\big]\Big\}
\nonumber\\&&{}\times\Gamma^{(0)}_2(\underline{z},\underline{x})\,,
\end{eqnarray}
where to our accuracy $\mathbb{H} = u \mathbb{H}^{(1)}$.

The CWI~(\ref{CWI0a}) for the two-point 1PI Green function takes the form
\begin{eqnarray}\label{CWI2}
0&=& \Big(K^{x_1}_{\bar n}(\widetilde \Delta)+K^{x_2}_{\bar n}(\widetilde \Delta)+2(n\bar n) Z\bar S_+
Z^{-1}\Big) \Gamma_2(\underline{z},\underline{x})
\nonumber\\&&{}
-\VEV{\delta_cS_R(\varphi)\,[\mathcal{O}(z_1,z_2)]\varphi(x_1)\varphi(x_2)}_{1PI}\,,
\end{eqnarray}
where (\ref{S+exp})
\begin{align}
  Z\bar S_+Z^{-1} = S_+^{(0)} - \epsilon(z_1+z_2) -\frac12 u[\mathbb{H}^{(1)},z_1+z_2]\,.
\end{align}
Substituting the expression  (\ref{DScorr}) into Eq.~(\ref{CWI2})
we can replace, to the required accuracy,
$\Gamma^{(0)}_2(\underline{z},\underline{x}) \to \Gamma_2(\underline{z},\underline{x})$ in the last term.
One sees then that the resulting contribution $\sim \gamma_\varphi$ to the CWI~(\ref{CWI2})
can be absorbed by modifying the parameter $\widetilde \Delta$ in
the conformal generators $K_{\bar n}(\widetilde \Delta)\to K_{\bar n}(\widetilde \Delta_\varphi)$
where
$$
\widetilde \Delta_\varphi = \widetilde \Delta-\gamma_\varphi
$$
is the (one-loop) shadow scaling dimension of the scalar field.
In this way we obtain
\begin{eqnarray}\label{CWI3}
0 &=& \Big\{K^{x_1}_{\bar n}(\widetilde \Delta_\varphi)+K^{x_2}_{\bar n}(\widetilde \Delta_\varphi)
+2(n\bar n)[\beta(u)\partial_u+ S_+] \Big\}
\nonumber\\&&{}\times\Gamma_2(\underline{z},\underline{x})+\mathcal{O}(u^2)\,,
\end{eqnarray}
where
\begin{align}\label{S+oneloop}
S_+=z_1^2\partial_{z_1}+z_2^2\partial_{z_2} + (z_1+z_2)\Big(\Delta+\frac12u \mathbb{H}^{(1)}\Big),
\end{align}
At the critical point, $u\to u_*$, $\beta(u_*)=0$, the contribution of the beta-function to Eq.~(\ref{CWI3}) vanishes
and we end up with the result for the deformation $\Delta S_+$ that has been quoted in Eq.~(\ref{DeltaS0+}).

The one-loop expression in (\ref{S+oneloop}) is the same for both scalar theories that we consider in this paper.
It is tempting to assume that this result can be generalized to all orders as
\begin{align}\label{S+conjecture}
S_+=z_1^2\partial_{z_1}+z_2^2\partial_{z_2} + (z_1+z_2)\Big(\Delta+\frac12\mathbb{H}(u_*)\Big)\,.
\end{align}
Indeed, the expression in Eq.~(\ref{S+conjecture})
obeys the necessary commutation relation $[S_+,S_-]=2S_0$ and its action on
the eigenfunctions $\Psi_{Nk}(z_1,z_2)$  takes the expected form
\begin{align}\label{Conj84}
S_+\Psi_{Nk}=\left(z_1^2\partial_{z_1}+z_2^2\partial_{z_2}+(z_1+z_2)\Delta_N\right)\Psi_{Nk}\,.
\end{align}
It can be shown that Eq.~(\ref{Conj84}) results in the form of a conformal operator proposed in
Ref.~\cite{Brodsky:1984xk}.

It turns out, however, that this (plausible) conjecture is  wrong.
We  have calculated the operator $S_+$ in the $\varphi^4$ theory at the order $\epsilon^2$ with the result
\begin{align}\label{S+NL}
S_+=&z_1^2\partial_{z_1}+z_2^2\partial_{z_2}+(z_1+z_2)\Big(\Delta+\frac12\mathbb{H}(u_*)\Big)
\notag\\
&+\frac14 u_*^2\, [\mathbb{H}^{(2)},z_1+z_2]+\mathcal{O}(\epsilon^3)\,,
\end{align}
where $\mathbb{H}^{(2)}$ is the two-loop evolution kernel,
Eq.~(\ref{Hexp}). Details of this calculation can be found in \ref{app:B}.

\section{Constraints for the evolution kernels}

 It is easy to see that translations along the $n$ direction,
 $(z_1,z_2)\mapsto (z_1+a,z_2+a)$, and scale transformations
 $(z_1,z_2)\mapsto (\lambda z_1,\lambda z_2)$ commute with the evolution
 operator~$\mathbb{H}$.
 The first property is an obvious consequence of Poincare invariance,
 and the second one is equivalent to the statement that only the operators of the
 same canonical dimension mix under renormalization. A generic integral operator satisfying these
 two restrictions can be represented in the form
\begin{align}\label{int-H-form}
[\mathbb{H} f](z_1,z_2)=\int d\alpha d\beta\, h(\alpha,\beta)
f(z_{12}^\alpha,z_{21}^\beta)\,,
\end{align}
where
\begin{align}
z_{12}^\alpha\equiv z_1\bar\alpha+z_2\alpha && \bar\alpha=1-\alpha\,,
\end{align}
and $h(\alpha,\beta)$ is a certain weight function. Note that the powers
$f(z_1,z_2) = (z_{1}-z_{2})^N$ are eigenfunctions of the evolution kernel $\mathbb{H}$,
and the corresponding eigenvalues
\begin{align}\label{hmoments}
\gamma_N=\int d\alpha d\beta\, h(\alpha,\beta)(1-\alpha-\beta)^N\,.
\end{align}
are nothing else as the anomalous dimensions, $\gamma_N$.

In general the function $h(\alpha,\beta)$ is a function of two variables. However, if $\mathbb{H}$ is an invariant
operator with respect to the canonical conformal transformations (\ref{S_0}), $[\mathbb{H}_{\rm inv },S_\alpha^{(0)}]=0$,
then it can be shown that the function $h(\alpha,\beta)$ takes the form~\cite{Braun:1999te}
\begin{align}\label{hinv}
h_{\rm inv }(\alpha,\beta)=(\bar\alpha\bar\beta)^{2j-2}\, \bar h\left(\frac{\alpha\beta}{\bar\alpha\bar\beta}\right)\,
\end{align}
and is effectively a function of one variable. This function can easily be reconstructed from its moments
(\ref{hmoments}), alias from the anomalous dimensions.

In the interacting theory $[\mathbb{H},S_\alpha^{(0)}]\slashed{=}0$ beyond the leading order.
As it was shown in the previous section one can define, however, three generators
$S_\alpha = S_{\alpha}^{(0)}+ \Delta S_{\alpha} $ (\ref{DeltaS}), (\ref{DeltaS0+})
which  satisfy the canonical $sl(2)$ commutation relations~(\ref{sl2-comm})
(for the theory at the critical coupling in non-integer dimensions).
The commutation relations impose certain self-consistency relations on the corrections
$\Delta S_{\alpha}$. Since the evolution kernel $\mathbb{H}(u_\ast)$ commutes with two of the generators, $S_-$ and $S_0$,
and since, as it is easy to see~\footnote{Indeed, $S_0^{(0)}$  counts the canonical
dimension on an object},
\begin{align}
  {}[S_0^{(0)},\Delta S_+]=\Delta S_+\,,
\end{align}
there are two such relations only:
\begin{align}\label{restr}
{}[S_+^{(0)},\Delta S_0]=&~[\Delta S_0, \Delta S_+]\,,
\notag\\
{}[\Delta S_+, S_-]=&~2\Delta S_0\,.
\end{align}
Taking into account that $\Delta S_0 = -\epsilon + (1/2)\mathbb{H}(u_*)$, see Eq.~(\ref{DeltaS0+}), the first relation
in (\ref{restr}) can be rewritten as
$[S_+^{(0)},\mathbb{H}(u_*)]=[\mathbb{H}(u_*), \Delta S_+]$. It
implies that the exact evolution kernel $\mathbb{H}(u_*)$ commutes with the full generator $S_+$.
The second relation provides a constraint on the possible deformation of $S_+$.

Writing $\Delta S_+$ as a power series in the critical coupling $u_\ast$
\begin{align}
\Delta S_+= \sum_{k=1}^\infty u_*^k
\Delta S_+^{(k)}\,
\end{align}
and expanding $[S_+,\mathbb{H}(u_\ast)]=0$ in powers of $u_*$ one obtains
\begin{align}\label{HE}
[S_+^{(0)},\mathbb{H}^{(1)}]=&~0\,,
\notag\\
[S_+^{(0)},\mathbb{H}^{(2)}]=&~[\mathbb{H}^{(1)},\Delta S_+^{(1)}]\,,
\notag\\
[S_+^{(0)},\mathbb{H}^{(3)}]=&~[\mathbb{H}^{(1)},\Delta S_+^{(2)}]+[\mathbb{H}^{(2)},\Delta S_+^{(1)}]\,,
\end{align}
etc. Note that the commutator of the canonical generator $S_+^{(0)}$ with the evolution kernel at order $\ell$
is given in terms of the evolution kernels $\mathbb{H}^{(k)}$ and the corrections to the generators
$\Delta S_+^{(k)}$ at one order less, $k \le \ell-1$.
The commutation relations  Eq.~(\ref{HE}) can be viewed as, essentially, inhomogeneous first-order
differential equations on the evolution kernels. Their solution determines $\mathbb{H}^{(k)}$
up to an $sl(2)$-invariant term, $[\mathbb{H}_{inv}^{(k)},S_\alpha^{(0)}]=0$, which can be restored
from the spectrum of the anomalous dimensions.

As discussed above, an evolution kernel $\mathbb{H}$ can be represented in the form of an integral
operator~(\ref{int-H-form}). The corresponding weight function $h(\alpha,\beta)$ can be split
in the $sl(2)$-invariant and non-invariant parts with respect to the canonical transformations (\ref{S_0})
\begin{align}
 h^{(k)}(\alpha,\beta) = h^{(k)}_{\rm inv}(\alpha,\beta)+\Delta h^{(k)}(\alpha,\beta)\,,
\label{hsplit}
\end{align}
cf. Eq.~(\ref{inv-noninv}). We will show that the non-invariant part
can be determined from the commutation relations and requires a $(k-1)$-loop calculation.
In turn, the invariant function $h^{(k)}_{{\rm inv}}(\alpha,\beta)$ takes the form (\ref{hinv})
and can easily be restored from the moments.
Let $\Delta\gamma^{(k)}_{N}$ be the eigenvalue of $\Delta\mathbb{H}^{(k)}$ on the function $(z_1-z_2)^N$,
i.e.
\begin{align}
 \Delta\gamma^{(k)}_{N} = \int d\alpha d\beta\, \Delta h^{(k)}(\alpha,\beta) (1-\alpha-\beta)^N\,.
\end{align}
Since the invariant function $h^{(k)}_{{\rm inv}}(\alpha,\beta)$ is effectively a function of one variable~(\ref{hinv})
it can be recovered inverting the equation for the  moments
\begin{align}
\int d\alpha d\beta\, h^{(k)}_{{\rm inv}}(\alpha,\beta) (1-\alpha-\beta)^N =  \gamma^{(k)}_{N} - \Delta\gamma^{(k)}_{N}
\end{align}
Determination of the anomalous dimensions $\gamma_N^{(k)}$ still requires evaluation of $k$-loop integrals.
It is much simpler, nevertheless, than  calculation of the full kernel $h^{(k)}(\alpha,\beta)$
alias the full anomalous dimension matrix at the same order.
In what follows we demonstrate the utility of this procedure on two examples.

\subsection{Three-loop evolution equations in the $\varphi^4$ theory}

Twist-two operators in the $O(n)$-symmetric $\varphi^4$ theory can be divided in three classes that
transform differently under rotations in the isotopic space: scalar (sc), symmetric and traceless (st) and antisymmetric (as)
\begin{eqnarray}
 \mathcal{O}_{(sc)}(z_1,z_2) &=& \varphi^a(z_1n)\varphi^a(z_2n)\,,
\notag\\
 \mathcal{O}_{(st)}^{ab}(z_1,z_2) &=&  [\varphi^a(z_1n)\varphi^b(z_2n) + (a\leftrightarrow b)] -\,\text{trace}\,,
\notag\\
 \mathcal{O}_{(as)}^{ab}(z_1,z_2) &=&  \varphi^a(z_1n)\varphi^b(z_2n) - (a\leftrightarrow b)\,,
\end{eqnarray}
respectively.
Anomalous dimensions for all these operators were calculated in Ref.~\cite{Kehrein:1995ia} at two loops,
and for the scalar operators in Ref.~\cite{Derkachov:1997pf} at four loops.
The anomalous dimensions of the symmetric traceless and the antisymmetric operators
can easily be derived from the expressions presented in Ref.~\cite{Derkachov:1997pf}
by taking into account appropriate isotopic factors. We collect below the anomalous
dimensions $\gamma_k$ ($k$ is the number of derivatives) to three-loop accuracy that is
relevant for this study.

For the scalar operators one obtains~\cite{Derkachov:1997pf}
\begin{align}
\gamma^{(sc)}_{k = 0}(u)=&(n+2)\biggl\{\frac{u}3-\frac{5u^2}{18}+\frac{u^3(5n+37)}{36}+\ldots\biggr\},
\notag\\
\gamma^{(sc)}_{k \ge 1}(u)=&(n+2)\biggl\{\frac{u^2}{18}\frac{(k-2)(k+3)}{k(k+1)}-\frac{2u^3(n+8)}{27 k(k+1)}
\notag\\ &
\times\left[
{S_1(k)}
+\frac{k^4+2k^3-39k^2-16k+12}{16k(k+1)}\right]\biggr\}
\notag\\ & +\ldots,
\end{align}
where  $S_1(k)=\sum_{m=1}^k 1/m=\psi(k+1)-\psi(1)$.
Note that the anomalous dimension $\gamma_{k=2}$ vanishes:
The corresponding operator is nothing but the energy momentum tensor of the scalar field
and it is conserved in quantum theory.

For the symmetric traceless operators we get
\begin{align}\label{}
\gamma_{k=0}^{(st)}(u)=&\,\frac{2u}{3}-\frac{u^2(n+10)}{18}
-\frac{u^3(5n^2-84n-444)}{216}
\nonumber\\ &+\ldots\,,
\nonumber\\
\gamma_{k\ge 1}^{(st)}(u)=&\frac{u^2}{9}\biggl[\frac{n+2}2-\frac{n+6}{k(k+1)}\biggr]-\!\frac{u^3}{54}\biggl[
\frac{(n+2)(n+8)}4
\notag\\
&+\frac{8(n+4)}{k(k+1)}\left(2S_1(k)-\frac{4k^2+2k-1}{k(k+1)}\right)
\notag\\
&-\frac{(2k^2-1)(n^2+6n+16)}{k^2(k+1)^2}\biggr]+
\ldots
\end{align}
and, finally, for the antisymmetric operators
\begin{align}
\gamma_{k\ge1}^{(as)}(u)=&(n+2)\biggl\{\frac{u^2}{9}\left[\frac12-\frac{1}{k(k+1)}\right]-\frac{u^3}{54}\biggl[
\frac{(n+8)}4
\notag\\
&+\frac{4}{k(k+1)}\left(2S_1(k) -\frac{4k^2+2k-1}{k(k+1)}\right)
\notag\\
&-\frac{(2k^2-1)(n+4)}{k^2(k+1)^2}\biggr]+
\ldots \biggr\}.
\end{align}
Note that the antisymmetric operator without derivatives $(k=0)$ does not exist.
These results are in agreement with the $1/n$-expansion of the anomalous dimensions~\cite{Derkachov:1997ch}.

We now proceed with the calculation of the evolution kernels.
As the first step one has to reconstruct the leading-order evolution kernel $\mathbb{H}^{(1)}$.
This operator commutes, cf. the first relation in~(\ref{HE}), with canonical $sl(2)$ generators
$S_\alpha^{(0)}$, $\alpha = 0,\pm$ and its spectrum is
\begin{align}\label{c-coef}
 \gamma^{(1)}_k =&c_n\,\delta_{k0}\,, &&
 c_n=\left\{\frac{n+2}3,\,\frac23,\,0\right\},
\end{align}
for the scalar, symmetric traceless, and antisymmetric operators, respectively.
The general form of the invariant operator is given by Eq.~(\ref{hinv}) where, for the case at hand,
$j=1/2$.  It is easy to convince oneself that the spectrum  $\gamma_k\sim \delta_{k0}$
corresponds to the choice $\bar h(\tau)\sim\delta(1-\tau)$.
We define a $sl(2)$ invariant operator~\cite{Derkachov:1995zr,Braun:2009vc}
\begin{align}\label{Hd}
{}[\mathcal{H}^{(d)}f](z_1,z_2)=&\int_0^1 \frac{d\alpha}{\bar\alpha}\int_0^{\bar\alpha}\frac{d\beta}{\bar\beta}
\delta\left(1-\frac{\alpha\beta}{\bar\alpha\bar\beta}\right)
 f(z_{12}^\alpha,z_{21}^\beta)
\notag\\
=&\int_0^1 d\alpha\,f(z_{12}^\alpha,z_{12}^\alpha)\,.
\end{align}
Obviously
$$
\mathcal{H}^{(d)}z_{12}^k=\delta_{k0} z_{12}^k\,,\qquad z_{12} \equiv z_1-z_2\,,
$$
so that the one-loop evolution kernel can be written as
\begin{align}\label{H1(4)}
\mathbb{H}^{(1)}=c_n\, \mathcal{H}^{(d)}\,,
\end{align}
where the coefficient $c_n$ depends on the symmetry of the operators, cf. Eq.~(\ref{c-coef}).

Note that the operator $\mathcal{H}^{(d)}$ is a $sl(2)$-invariant projector.
Indeed, one can easily check that
$$(\mathcal{H}^{(d)})^2=\mathcal{H}^{(d)}\,,\qquad\mathcal{H}^{(d)}(1-\mathcal{H}^{(d)})=0.$$
For later use we define the operator
\begin{align}
 \mathrm{\Pi}_0 = 1-\mathcal{H}^{(d)}
\label{Pi0}
\end{align}
which is also a projector. \

As the next step, we  calculate the non-invariant part of the two-loop
kernel~$\mathbb{H}^{(2)}$.  Re-expanding $\epsilon =(4-d)/2$ in terms of the critical
coupling
\begin{align}
\epsilon=\frac{n+8}{6} u_*-\frac{3n+14}{6}u_*^2+\mathcal{O}(u_*^3)\,,
\end{align}
cf. Eq.~(\ref{ust4}), one obtains the one-loop deformation of the generator of special
conformal transformations:
\begin{align}
\Delta S_+^{(1)}=(z_1+z_2)\biggl[-\frac16 {(n+8)}+\frac12\mathbb{H}^{(1)}\biggr].
\end{align}
Using this expression, the second commutator relation in Eq.~(\ref{HE}) takes the form
\begin{eqnarray}\label{HE2}
{}[S_+^{(0)},\mathbb{H}^{(2)}]&=&[\mathbb{H}^{(1)},{z_1+z_2}]\left\{-\frac16 {(n+8)}+\frac12\mathbb{H}^{(1)}\right\}
\notag\\
&=&\frac{c_n}{2}[\mathcal{H}^{(d)},z_1+z_2]\left\{c_n-\frac13 {(n+8)}\right\},
\end{eqnarray}
where in the second line we have taken into account that
$[\mathcal{H}^{(d)},z_1+z_2]\mathcal{H}^{(d)}=[\mathcal{H}^{(d)},z_1+z_2]$.
The remaining commutator on the r.h.s. of (\ref{HE2}) can be written as
\begin{eqnarray}
{}[\mathcal{H}^{(d)},z_1+z_2]&=&z_{12}\widetilde{\mathcal{H}}^{(d)}\,,
\nonumber\\
{}[\widetilde{\mathcal{H}}^{(d)}  f](z_1,z_2)&=&\int_0^1 d\alpha\,(\bar\alpha-\alpha)f(z_{12}^\alpha,z_{12}^\alpha)\,,
\label{Htilde}
\end{eqnarray}
which is easy to verify using the explicit expression in Eq.~(\ref{Hd}).

Eq.~(\ref{HE2}) can be viewed as an equation on $\mathbb{H}^{(2)}$. We look for
the solution as the sum (\ref{inv-noninv})
\begin{align}
  \mathbb{H}^{(2)}= \mathbb{H}^{(2)}_{inv}+ \Delta\mathbb{H}^{(2)}
\label{H2split}
\end{align}
such that $\mathbb{H}^{(2)}_{inv}$ is a solution of the homogeneous equation $[\mathbb{H}^{(k)}_{\rm inv},S^{(0)}_{\pm,0}] = 0$.

It is easy to check that for an operator that has the structure
\begin{align}\label{VV}
[\mathbb{H}f](z_1,z_2)=\int_0^1 d\alpha\,\upsilon(\alpha)\,f(z_{12}^\alpha,z_{12}^\alpha)\,
\end{align}
the  commutator with $S_{+}^{(0)}$ equals to
\begin{align}\label{VVV}
[[S_{+}^{(0)},\mathbb{H}]f](z_1,z_2)=
z_{12}\!\int_0^1 d\alpha\,\alpha\bar\alpha\, \partial_\alpha\upsilon(\alpha)
f(z_{12}^\alpha,z_{12}^\alpha)\,.
\end{align}
It follows from Eqs.~(\ref{HE2}) and
(\ref{Htilde})  that $\Delta\mathbb{H}^{(2)}$ corresponds to the weight function
\begin{align}
\upsilon(\alpha)=\tilde c_n\, (\ln\alpha+\ln\bar\alpha)\,,
\end{align}
where
\begin{align}\label{}
\tilde c_n =\frac{c_n}{6}\left[3c_n- {(n+8)}\right]=\Big\{-\frac{n\!+\!2}3,-\frac{n\!+\!6}9,0
\Big\}
\end{align}
for the scalar, symmetric traceless and antisymmetric operators, respectively.
Thus
\begin{align}\label{deltaH2}
\Delta\mathbb{H}^{(2)} = \tilde c_n\, \mathcal{V}^{(d,1)}\,,
\end{align}
where
\begin{align}\label{Hd1}
{}[\mathcal{V}^{(d,1)}f](z_1,z_2)=&\int_0^1d\alpha\, \ln\alpha\bar\alpha f(z_{12}^\alpha,z_{12}^\alpha)\,.
\end{align}
A straightforward calculation yields
\begin{align}
{}[\Delta\mathbb{H}^{(2)}z_{12}^N](z_1,z_2)= -4 \tilde c_n\,\delta_{N0} \, z^N_{12} \equiv \Delta\gamma^{(2)}_N \, z^N_{12}\,.
\end{align}
As the last step, the invariant kernel $\mathbb{H}^{(2)}_{inv}$ (\ref{H2split}) can be reconstructed from the
known two-loop anomalous dimensions $\gamma^{(2)}_N$ by inverting the equation for the moments
\begin{align}
{}[\mathbb{H}^{(2)}_{inv}z_{12}^N](z_1,z_2) = (\gamma^{(2)}_N - \Delta\gamma^{(2)}_N) \, z^N_{12}\,.
\end{align}
One gets after some algebra%
\footnote{Here and below we use a generic notation $\mathcal{H}^{(a)}$ for the $sl(2)$-invariant and
$\mathcal{V}^{(a)}$ for the $sl(2)$-breaking kernels}
\begin{align}\label{invH2}
\mathbb{H}^{sc,(2)}=&-\frac{n+2}3\left[3\mathcal{H}^{(d)}+\mathcal{H}^{(+)}\mathrm{\Pi}_0+\mathcal{V}^{(d,1)}-\frac16\right],
\notag\\
\mathbb{H}^{st,(2)}=&-\frac{n+6}9\left[3\mathcal{H}^{(d)}+\mathcal{H}^{(+)}\mathrm{\Pi}_0+\mathcal{V}^{(d,1)}\right]+\frac{n+2}{18}\,,
\notag\\
\mathbb{H}^{as,(2)}=&-\frac{n+2}9\left[\mathcal{H}^{(+)}-\frac12\right],
\end{align}
where we have introduced a new invariant kernel $\mathcal{H}^{(+)}$ that corresponds to the choice
$\bar h(\tau)=(1-\tau)^{-1}$ in Eq.~(\ref{hinv}):
\begin{align}
{}[\mathcal{H}^{(+)}f](z_1,z_2)=&\int_0^1d\alpha\int_0^{\bar\alpha}\frac{d\beta}{1-\alpha-\beta}
{f(z_{12}^\alpha,z_{21}^\beta)}\,.
\end{align}
Note that $\mathcal{H}^{(+)}$ is only well defined on the space of functions that vanish at $z_1\to z_2$.
It can be checked that the projector $\mathrm{\Pi}_0$ eliminates a constant term at $z_1\to z_2$
from any function, so that the product $\mathcal{H}^{(+)}\mathrm{\Pi}_0$ is always well defined.
We have removed the projector $\mathcal{H}^{(+)}\mathrm{\Pi}_0 \to \mathcal{H}^{(+)}$
in the last expression in Eq.~(\ref{invH2}) (for antisymmetric operators) since the relevant functions
are in this case antisymmetric under the permutations $z_1\leftrightarrow z_2$.

Proceeding in the same way one can derive the three-loop evolution kernels.
The second-order correction to the generator $S_+^{(0)}$ takes the form
\begin{align}
\Delta S_+^{(2)}=(z_1+z_2)\frac{3n+14}{6}+\frac14\big\{z_1+z_2,\mathbb{H}^{(2)}\big\}\,,
\end{align}
where $\{\ast,\ast\}$ stands for the anticommutator.
Taking into account that
\begin{align}
 \mathbb{H}^{(1)}[\mathbb{H}^{(2)},z_1+z_2]=  c_n \, \mathcal{H}^{(d)} [\mathbb{H}^{(2)},z_1+z_2] = 0
\end{align}
one obtains an equation for $\mathbb{H}^{(3)}$:
\begin{eqnarray}\label{SS3}
[S_+^{(0)},\mathbb{H}^{(3)}]&=&\frac16[({3n+14})\mathbb{H}^{(1)}-
({n+8})\mathbb{H}^{(2)}\!,z_1\!+\!z_2]
\nonumber\\
&+&\frac14[\mathbb{H}^{(2)}\!,z_1\!+\!z_2]\mathbb{H}^{(1)}
+\frac12[\mathbb{H}^{(1)}\!,z_1\!+\!z_2]\mathbb{H}^{(2)}\!.
\nonumber\\
\end{eqnarray}
Let us consider the antisymmetric operators at first.
This case is simpler because $\mathbb{H}^{as(1)}=0$ so that
one gets
\begin{align}\label{S3asym}
[S_+^{(0)},\mathbb{H}^{as(3)}]=\varkappa_n\, [\mathcal{H}^{+},z_1+z_2] = \varkappa_n z_{12}\widetilde{\mathcal{H}}^{+}\,,
\end{align}
where $$\varkappa_n={(n+2)(n+8)}/{54}$$
and
\begin{align}
{}[\widetilde{\mathcal{H}}^{(+)}f](z_1,z_2)=
\int_0^1d\alpha\int_0^{\bar\alpha}d\beta \frac{\beta-\alpha}{1-\alpha-\beta}f(z_{12}^\alpha,z_{21}^\beta)\,.
\label{tildeHplus}
\end{align}
The commutator of $S_+^{(0)}$ with an integral operator that has a generic structure (\ref{int-H-form})
\begin{align}
[\mathbb{H} f](z_1,z_2)= \int_0^1d\alpha \int_0^{\bar\alpha}d\beta\, h(\alpha,\beta)
f(z_{12}^\alpha,z_{21}^\beta)\,
\label{hform}
\end{align}
can be written as
\begin{eqnarray}
[[S_+^{(0)},\mathbb{H}]f](z_1z_2) &=&
z_{12}\int_0^1\!d\alpha\!\int_0^{\bar\alpha }\!\!d\beta f(z_{12}^\alpha,z_{21}^\beta)
\nonumber\\&& {}\hspace*{-0.8cm}
\times\big[\alpha\bar\alpha\partial_\alpha-\beta\bar\beta\partial_\beta+\beta-\alpha\big]h(\alpha,\beta)\,.
\end{eqnarray}
Looking for the solution of Eq.~(\ref{S3asym}) in this form one obtains
\begin{align}
h(\alpha,\beta)=\varkappa_n \frac{\ln(1\!-\!\alpha\!-\!\beta)}{1\!-\!\alpha\!-\!\beta}
+ \frac{1}{\bar\alpha\bar\beta} \bar h\left(\frac{\alpha\beta}{\bar\alpha\bar\beta}\right),
\end{align}
where the function $\bar h$ is arbitrary (a solution of the homogeneous equation).
It corresponds to the invariant kernel.
Thus
\begin{align}
  \Delta\mathbb{H}^{as(3)} = \varkappa_n \mathcal{V}^{(+,1)}
\end{align}
with
\begin{align}
\mathcal{V}^{(+,1)}=
\int_0^1d\alpha\int_0^{\bar\alpha }d\beta\, \frac{\ln(1\!-\!\alpha\!-\!\beta)}{1\!-\!\alpha\!-\!\beta} f(z_{12}^\alpha,z_{21}^\beta)\,.
\label{Vplus1}
\end{align}
As above, the $sl(2)$-invariant contribution can be restored from the known anomalous dimensions.
We obtain after some algebra
\begin{equation}
\mathbb{H}^{as(3)}=\varkappa_n\biggl[
\mathcal{V}^{(+,1)}
+\frac{4}{n+8}\mathcal{H}^{(1)}+2\frac{n+12}{n+8}\mathcal{H}^{(+)} -\frac14\biggr],
\label{Has3}
\end{equation}
where
\begin{align}
\mathcal{H}^{(1)}f =
\int_0^1d\alpha\int_0^{\bar\alpha }d\beta\,
\frac{1}{1\!-\!\alpha\!-\!\beta}\ln\left(\frac{\alpha\beta}{\bar\alpha\bar\beta}\right) f(z_{12}^\alpha,z_{21}^\beta)\,.
\end{align}
The kernels in Eq.~(\ref{Has3}) have the following eigenvalues on $z_{12}^k$:
\begin{eqnarray}
 \mathcal{H}^{(+)} z_{12}^k&=& \frac{1}{k(k+1)}\,z_{12}^k\,,
\nonumber\\
 \mathcal{H}^{(1)}z_{12}^k &=& -\frac{2S_1(k)}{k(k+1)}\,z_{12}^k\,,
\nonumber\\
 \mathcal{V}^{(+,1)}z_{12}^k &=& -\frac{2k+1}{k^2(k+1)^2}\,z_{12}^k\,.
\end{eqnarray}

\begin{table}
\caption{The expansion coefficients $r_k$ (\ref{Hrk}). The upper entries corresponds to
the scalar operators and the lower ones to the symmetric traceless operators.}
\label{tab:rk-coef}       
\begin{tabular}{c|cccc}
\hline\noalign{\smallskip}
& $r_1$ & $r_2$ & $r_3$  \\
\noalign{\smallskip}\hline\noalign{\smallskip}
{sc} & $ \frac{(n+2)(11n+100)}{54}$& $\frac{(n+2)(n+14)}{36}$ &$\frac{(n+2)(n+8)}{18}$ \\
\\
{st} & $\frac{3n^2+56n+200}{54}$& $\frac{(n+6)(n+7)}{54}$ & $\frac{(n+6)(n+8)}{54}$ \\
\noalign{\smallskip}\hline
\end{tabular}
\end{table}

The calculation of the three-loop evolution kernel for the scalar and symmetric traceless operators goes along the same lines
so that we will only sketch the main steps.
For the operators of the type
\begin{align}
{}[\mathcal{H}(w) f](z_1,z_2)=\int_0^1 d\alpha \,w(\alpha)\, f(z_{12}^\alpha,z_{12}^\alpha)
\end{align}
the following identity holds:
\begin{align}\label{HwHw}
\mathcal{H}(w_1)\mathcal{H}(w_2)=\left(\int_0^1d\alpha\, w_2(\alpha)\right)\mathcal{H}(w_1)\,
\end{align}
which appears to be
quite useful.

The commutator involving $\mathbb{H}^{(2)}$ on the r.h.s. of (\ref{SS3}) can be written as
\begin{align}
{}[\mathbb{H}^{(2)},z_1\!+\!z_2]=z_{12}\xi_n\left[3\widetilde{\mathcal{H}}^{(d)}\!+
\widetilde{\mathcal{H}}^{(+)}\mathrm{\Pi}_0+\!\widetilde{\mathcal{H}}^{(d,1)}\right]
\end{align}
where
\begin{align}
\xi_n=\big\{-({n+2})/{3}, -({n+6})/{9}\big\}
\end{align}
for the scalar and symmetric traceless operators, respectively.
The kernels $\widetilde{\mathcal{H}}^{(d)}$ and $\widetilde{\mathcal{H}}^{(+)}$
are defined above in Eqs.~(\ref{Htilde}), (\ref{tildeHplus}) and
\begin{align}
[\widetilde{\mathcal{H}}^{(d,1)}f](z_1,z_2)=\int_0^1d\alpha(\bar\alpha\ln\alpha-\alpha\ln\bar\alpha)
f(z_{12}^\alpha,z_{12}^\alpha).
\end{align}
A straightforward calculation yields:
\begin{align}\label{Hrk}
[S_+^{(0)}\!,\mathbb{H}^{(3)}]=\!z_{12}\!\left[r_1\widetilde{\mathcal{H}}^{(d)}\!+r_2\widetilde{\mathcal{H}}^{(d,1)}+
r_3\widetilde{\mathcal{H}}^{(+)}\mathrm{\Pi}_0\right]\!.
\end{align}

\begin{table}[t]
\caption{The expansion coefficients $p_k$ (\ref{Hinv3}). The upper entries corresponds to
the scalar operators and the lower ones to the symmetric traceless operators.}
\label{tab:pk-coef}       
\begin{tabular}{c|cccc}
\hline\noalign{\smallskip}
& $p_1$ & $p_2$ & $p_3$  \\
\noalign{\smallskip}\hline\noalign{\smallskip}
{\rm sc} & $ \frac{(n+2)(107n+862)}{216}$& $\frac{5(n+2)(n+8)}{27}$ &$\frac{(n+2)(n+8)}{27}$ \\
\\
{\rm st} & $\frac{6n^2+219n+862}{108}$& $\frac{n^2+22n+80}{27}$ & $\frac{4(n+4)}{27}$ \\
\noalign{\smallskip}\hline
\end{tabular}
\end{table}

\noindent
The expressions for the coefficients $r_1, r_2, r_3 $ are collected in Table~\ref{tab:rk-coef}.
In this way we obtain for the non-invariant part of the kernel
\begin{align}
\Delta\mathbb{H}^{(3)}=r_1{\mathcal{V}}^{(d,1)}+r_2{\mathcal{V}}^{(d,2)}+
r_3{\mathcal{V}}^{(+,1)}\mathrm{\Pi}_0\,.
\label{DeltaH3}
\end{align}
The operators ${\mathcal{V}}^{(d,1)}$ and ${\mathcal{V}}^{(+,1)}$ are defined in Eqs.~(\ref{Hd1}) and (\ref{Vplus1}), respectively.
The new contribution $\mathcal{V}^{(d,2)}$ comes in play as solution to the equation
$[S_+^{(0)},\mathcal{H}^{(d,2)}]=z_{12}\widetilde{\mathcal{H}}^{(d,1)}$ and has the form
\begin{align}
[\mathcal{V}^{(d,2)}f](z_1,z_2)=\frac12\int_0^1\! d\alpha\,(\ln^2\alpha+\ln^2\bar\alpha)f(z_{12}^\alpha,z_{12}^\alpha).
\end{align}

The remaining invariant part of the kernel $\mathbb{H}^{(3)}$ can be restored from the anomalous
dimensions.  The result can be written as follows
\begin{align}\label{Hinv3}
\mathbb{H}^{(3)}_{inv}=p_1\mathcal{H}^{(d)}+p_2\mathcal{H}^{(+)}\mathrm{\Pi}_0+p_3\mathcal{H}^{(1)}+
2\gamma_{\varphi}^{(3)}\,,
\end{align}
where $\gamma_{\varphi}^{(3)}=-{(n+2)(n+8)}/{432}$ is the corresponding coefficient in  the anomalous dimension of the scalar field (\ref{rgf-4})
and the coefficients $p_k$ are given in Table~\ref{tab:pk-coef}.
The total kernel $\mathbb{H}^{(3)}$ is given by the sum of the expressions in Eqs.~(\ref{DeltaH3}) and (\ref{Hinv3}),
$\mathbb{H}^{(3)}=\mathbb{H}^{(3)}_{inv}+\Delta\mathbb{H}^{(3)}$.

To summarize, making use of the {\it exact} conformal invariance of the scalar theory at the critical
coupling in $d=4-2\epsilon$ dimensions we have been able to restore the complete three-loop evolution kernels
(alias the full anomalous dimension matrix)
at arbitrary coupling $u$ using three-loop anomalous dimensions as input. The required calculation is mainly algebraic.
The only place where Feynman diagrams appear is the calculation of the deformation of the $S_+$ generator.
This calculation is, however, considerably simpler as compared to a direct evaluation of the three-loop  evolution kernels.

\subsection{Two-loop evolution equations in the $\varphi^3$ theory}

We use this example to discuss a somewhat different technique that is based on the representation of $sl(2)$
invariant kernels in terms of the Casimir operators~\cite{Bukhvostov:1985rn}.
To start with, we need to classify the existing twist-2 operators $\mathcal{O}^{ab}(z_1,z_2)$ according to the
irreducible representations of the isotopic $su(n)$ group.
We define
\begin{align}\label{Pj}
\mathcal{O}^{ab}_{j}(z_1,z_2)=(P_j)_{a'b'}^{ab}\mathcal{O}^{a'b'}(z_1,z_2),
\end{align}
where $P_j$, $j=1,\ldots,7$, are projectors onto the seven
irreducible representations
in the tensor product of two adjoint representations. Explicit expressions are given in~\ref{app:C}.
Operators corresponding to different representations do not mix under renormalization and can be considered
separately.

The one-loop evolution kernel (anomalous dimensions) for all operators except $\mathcal{O}_{j=3}$
is determined by the second diagram in Fig.~\ref{fig:1}.
For the case of $\mathcal{O}_{j=3}$ there is an additional contribution
corresponding to the transition $\varphi^{a}\varphi^{b}\to d^{abc}\partial^2\varphi^c$. This extra term
vanishes  for other operators thanks to the isotopic projector.
Although it does not present any particular complication for our analysis,
for simplicity we do not consider $\mathcal{O}_{j=3}$ in what follows.

The one-loop evolution kernel corresponding to the diagram in Fig.~\ref{fig:1} takes the form
\begin{align}
\mathbb{H}^{(1)}=-2\lambda_j\,\mathcal{H}+2\gamma^{(1)}_\varphi\,,
\label{3H11}
\end{align}
where $\lambda_j$ are numbers that depend on the representation and the rank of the group.
They are collected in Eq.~(\ref{lambdas}) in \ref{app:C}.
The operator $\mathcal{H}$ is defined as
\begin{align}
[\mathcal{H}f](z_1,z_2)=\int_0^1d\alpha\int_0^{\bar\alpha}d \beta f(z_{12}^\alpha,z_{21}^\beta)\,.
\end{align}
$\mathcal{H}$ commutes with the canonical $sl(2)$ generators $S^{(0)}_\alpha$
and has the following eigenvalues
\begin{align}
\mathcal{H} z_{12}^k = E_k z_{12}^k\,, &&  E_k=\frac1{(k+1)(k+2)}\,,
\label{Ek}
\end{align}
so that the one-loop anomalous dimensions of the twist-two operators are
equal to
\begin{align}
  \gamma^{(1)}_{j,k} =  -2\lambda_j\,E_k+2\gamma^{(1)}_\varphi\,, \qquad j \slashed{=}3\,,
\end{align}
where $k$ is the number of derivatives.

Comparing (\ref{Ek}) with the spectrum of the quadratic Casimir operator,
\begin{align}
&\mathbb{C}^{(0)}_2~=~S_+^{(0)}S_-^{(0)}+S_0^{(0)}(S_0^{(0)}-1)=-\partial_1\partial_2 z_{12}^2,
\notag\\
&\mathbb{C}^{(0)}_2z_{12}^k~=~(k+1)(k+2)z_{12}^k\,,
\end{align}
we conclude that the operator $\mathcal{H}$ is nothing else as the inverse of $\mathbb{C}^{(0)}_2$,
\begin{align}
\mathcal{H}=(\mathbb{C}^{(0)}_2)^{-1}.
\label{3H12}
\end{align}

The {complete} evolution kernel at the critical point, $\mathbb{H}(u_*)$,
commutes with the deformed generators $S_\alpha$ and hence is a function of the
complete (deformed) Casimir operator,
\begin{align}
\mathbb{H}_j(u_*)=h_j(\mathbb{C}_2)\,,
&&\mathbb{C}_2=S_+S_-+S_0(S_0-1)\,,
\end{align}
where the subscript $j$ enumerates the isotopic structures~(\ref{Pj}).

The function $h_j(x)$ has a perturbative expansion
\begin{align}
h_j(x)=&~u_*\, h_j^{(1)}(x)+u_*^2 \,h_j^{(2)}(x)+\ldots\,
\end{align}
and the leading contribution $h_j^{(1)}(x)$ is uniquely fixed by one-loop result
(\ref{3H11}), (\ref{3H12}), alias by the one-loop anomalous dimensions:
\begin{align}
h^{(1)}_j(x)= - 2 \lambda_j \frac{1}{x} +  2 \gamma_\varphi^{(1)}.
\end{align}
Expanding the Casimir operator $\mathbb{C}_2$
\begin{align}
\mathbb{C}_2=&~\mathbb{C}_2^{(0)}+u_* \,\mathbb{C}_2^{(1)}+\ldots
\end{align}
one gets the following expression for the evolution kernel $\mathbb{H}(u_*)$
to the $\mathcal{O}(u_\ast^3)$ accuracy:
\begin{eqnarray}\label{Hhh}
\mathbb{H}(u_*)&=&u_* h^{(1)}\left(\mathbb{C}_2^{(0)}\!+\!u_* \,\mathbb{C}_2^{(1)}\right)+
u_*^2\, h^{(2)}\,\Big(\mathbb{C}_2^{(0)}\Big).
\end{eqnarray}
Similar to the case of the $\varphi^4$ theory considered in the previous section,
our strategy here is to look for the the two-loop kernels $\mathbb{H}_j^{(2)}$ in the form
\begin{align}
 \mathbb{H}_j^{(2)} = \mathbb{H}_{j,inv}^{(2)} +  \Delta\mathbb{H}_j^{(2)}\,,
\end{align}
where the two terms correspond to the $sl(2)$-invariant and non-invariant contributions, respectively.
Calculation of the non-invariant kernel $\Delta\mathbb{H}_j^{(2)}$ is the main task, after which
the invariant kernel can easily be reconstructed from the spectrum of two-loop anomalous dimensions.

The last term in Eq.~(\ref{Hhh}) is obviously invariant under canonical $sl(2)$ transformations
so that one does not need to know $h^{(2)}(x)$;
$\Delta\mathbb{H}_j^{(2)}$ arises exclusively from the first term. Using Eq.~(\ref{3H12}) yields
\begin{align}\label{Cexp}
\left(\mathbb{C}_2^{(0)}+u_* \,\mathbb{C}_2^{(1)}\right)^{-1}=\mathcal{H}-u_* \mathcal{H}\mathbb{C}_2^{(1)}\mathcal{H}
+\mathcal{O}(u_*^2)\,.
\end{align}
Next, making use of explicit expressions for the deformed generators, Eqs.~(\ref{S_0}), (\ref{DeltaS0+}),
and writing
\begin{align}
\epsilon-\gamma_\varphi=u_*\kappa+\mathcal{O}(u_*^2),
&& \kappa=(16-n^2)/{3n}\,,
\end{align}
we obtain a correction to the Casimir operator
\begin{align}\label{C21}
\mathbb{C}_2^{(1)}=-\Big(\partial_1 z_{12}+\partial_2z_{21}+1\Big)\left(\kappa+\lambda_j \mathcal{H}\right).
\end{align}
Since we are interested here in the  $sl(2)$-breaking contributions to Eq.~(\ref{Cexp}) only, any
$sl(2)$-invariant terms in $\mathbb{C}_2^{(1)}$ can be dropped.
It is convenient to rewrite Eq.~(\ref{C21}) as follows:
\begin{align}
\mathbb{C}_2^{(1)}=-
\big(S_{12}+S_{21}\big)\left(\kappa+\lambda_j \mathcal{H}\right)
+\ldots,
\end{align}
where the ellipses stand for the $sl(2)$-invariant contributions and
\begin{align}
S_{12}={z_{12}^{-1}}\partial_1 z^2_{12}\,, && S_{21}={z_{21}^{-1}}\partial_2z^2_{21}\,
\end{align}
are intertwining operators%
\footnote{These relations follow readily from the intertwining relations for the generators
$$
z_{12}\,\left(S^{(j)}_{1,\alpha}+S^{(j)}_{2,\alpha}\right)=\left(S^{(j-1/2)}_{1,\alpha}+S^{(j-1/2)}_{2,\alpha}\right)\,z_{12}
$$
and $\partial_z S^{j=0}=S^{j=1}\partial_z$. Both are easy to check.}
\begin{align}
S_{12} T^{j=1}\otimes T^{j=1}=T^{j=3/2}\otimes
T^{j=1/2}S_{12}\,,\notag\\
S_{21} T^{j=1}\otimes T^{j=1}=T^{j=1/2}\otimes
T^{j=3/2}S_{21}\,.
\end{align}
Thus we have to evaluate the following expression:
\begin{align}
\mathcal{H}\mathbb{C}_2^{(1)}\mathcal{H}=
-\mathcal{H}\big(S_{12}+S_{21}\big)\left(\kappa+\lambda_j \mathcal{H}\right)\mathcal{H}\,.
\end{align}
To this end, the following technique proves to be very efficient.

As the first step, consider the operators
\begin{align}
\mathbb{W}_1=S_{12}\mathcal{H}\,, && \mathbb{W}_2 = S_{12}\mathcal{H}^2.
\end{align}
They are, both, $sl(2)$-invariant operators that act on
$T^{j=1}\otimes T^{j=1}\to T^{j=3/2}\otimes T^{j=1/2}$.
The general form of such an operator is given by the following expression (see e.g.~\cite{Braun:2009vc}):
\begin{align}
[\mathbb{W}f](z_1,z_2)=\int d\alpha d\beta\, \frac{\beta}{\bar\beta}\,
w\left(\frac{\alpha\beta}{\bar\alpha\bar\beta}\right) f(z_{12}^\alpha, z_{21}^\beta)
\end{align}
and the kernel $w(\tau)$ is completely determined by the spectrum
\begin{align}
 \mathbb{W}z_{12}^k = w_k z_{12}^k.
\end{align}
By a direct calculation one finds for the operators in question
\begin{align}
 w_{1,k} = \frac{1}{k+1}\,, && w_{2,k} =\frac{1}{(k+2)(k+1)^2}\,.
\end{align}
It is easy to check that the corresponding kernels
are%
\begin{align}
 w_{1,k} \mapsto w_1(\tau)=\delta(\tau)\,, && w_{2,k} \mapsto w_2(\tau)=1\,,
\end{align}
so that we obtain
\begin{align}
{}[\mathbb{W}_1f](z_1,z_2)=&\int_0^1 d\beta \ f(z_{1}, z_{21}^\beta)\,,
\notag\\
{}[\mathbb{W}_2f](z_1,z_2)=&\int_0^1 d\alpha \int_0^{\bar\alpha}d\beta\, \frac{\beta}{\bar\beta}\, f(z_{12}^\alpha, z_{21}^\beta).
\end{align}
Next, since $S_{21}=P_{12}S_{12}P_{21}$ where $P_{12}$ is the permutation operator $z_1\leftrightarrow z_2$,
and $[\mathcal{H},P_{12}]=0$, we can write
\begin{align}
\mathcal{H}\mathbb{C}_2^{(1)}\mathcal{H}=-\kappa\, \mathbb{U}_1-\lambda_j\,\mathbb{U}_2\,,
\end{align}
where
\begin{align}
\mathbb{U}_k=\mathcal{H}\,\mathbb{W}_k+P_{12}\,\mathcal{H}\,\mathbb{W}_k\, P_{12}\,.
\end{align}
One obtains after some algebra
\begin{align}
{}[\mathbb{U}_k f](z_1,z_2)=\int_{0}^1\!d\alpha\int_0^{\bar\beta}\!\!d\beta\, u_k(\alpha,\beta)f(z_{12}^\alpha, z_{21}^\beta)\,,
\end{align}
where%
\footnote{The full expression for $u_2$ involves the dilogarithm function and
can be brought to the the form in Eq.~(\ref{u12}) using the pentagon identity for $\Li_2$.}
\begin{align}\label{u12}
u_1(\alpha,\beta)=&-\ln(1\!-\!\alpha\!-\!\beta)+\ldots\,,
\\
u_2(\alpha,\beta)=&\frac12\Big[\ln^2(1\!-\!\alpha\!-\!\beta)-\ln^2\bar\alpha-\ln^2\bar\beta\Big]+\ldots.
\notag
\end{align}
The ellipses stand for contributions that are functions of the invariant (conformal) ratio
$ r =\alpha\beta/(\bar\alpha\bar\beta)$.
They give rise to $sl(2)$-invariant contributions to $\mathbb{C}_2^{(1)}$ and can be dropped in the
present context.

Collecting everything we obtain for the $sl(2)$-breaking part of the evolution kernel
\begin{eqnarray}
\lefteqn{ [\Delta\mathbb{H}^{(2)}f](z_1,z_2)=}
\nonumber\\&=&
\lambda_j\int_0^1d\alpha\!\int_{0}^{\bar\alpha}\!\!d\beta \Big[2\kappa \ln(1\!-\!\alpha\!-\!\beta)
- \lambda_j \ln^2(1\!-\!\alpha\!-\!\beta)
\nonumber\\&& \hspace*{2cm}
+  \lambda_j \ln^2\bar\alpha + \lambda_j \ln^2\bar\beta\Big]f(z_{12}^\alpha,z_{21}^\beta).
\end{eqnarray}
We have checked that this expression coincides with the result
of the direct calculation of the relevant Feynman diagrams.

The invariant part of the kernel (see \ref{app:two-loop}) has the form
\begin{align}
{}[\mathbb{H}_{inv}^{(2)}f](z_1,z_2)=\!\int_0^1\!d\alpha\int_{0}^{\bar\alpha}\!\!d\beta\, w(\alpha,\beta)
f(z_{12}^\alpha, z_{21}^\beta),
\end{align}
where
\begin{align*}
w(\alpha,\beta)=&
4\left[\frac12\nu_j-\lambda_j^2-\lambda_j\frac{n^2-4}{24n}\right]\ln\left(1-\frac{\alpha\beta}{\bar\alpha\bar\beta}\right)
\notag\\
&+\frac{\lambda_j}{3n}\!\left[92-5n^2-(n^2-16)\ln\left(\frac{\alpha\beta}{\bar\alpha\bar\beta}\right)\right].
\end{align*}
Explicit expressions for the  isotopic $su(n)$ factors $\lambda_j$ and  $\nu_j$ are given
in Eqs.~(\ref{lambdas}) and (\ref{nus}), respectively. It can be checked that the
anomalous dimensions $\gamma_{j=1,k=2}$ and $\gamma_{j=2,k=1}$ vanish as they should,
since the
corresponding operators are the energy momentum tensor and isotopic current, respectively.

\section{Summary}
We have studied implications of exact conformal invariance of scalar quantum field
theories at the critical point in non-integer dimensions for the evolution kernels of the light-ray operators.
The possibility to make this connection is based on the observation that in $\text{MS}$-like schemes the evolution kernels (anomalous dimensions)
do not depend on the space-time dimension. Thus all expressions derived in the $d$-dimensional
(conformal) theory remain exactly the same for the theory in integer dimensions.
We demonstrate that all conformal symmetry constraints for the twist-two light-ray operators are encoded
in the form of the generators of the so-called collinear $sl(2)$ subgroup.
Two of them, $S_-$ and $S_0$, can be fixed at all loops in terms of the evolution kernel,
while the generator of special conformal transformations, $S_+$, receives  nontrivial
corrections which can only be calculated order by order in perturbation theory.
Provided that the generator $S_+$ is known at the $\ell-1$ loop order, one can determine the evolution
kernel to the $\ell$-loop accuracy up to terms that are invariant with respect to the tree-level generators.
The invariant parts can eventually be restored from the anomalous dimensions.
This procedure is advantageous as compared to a direct calculation
because the calculation of the anomalous dimensions is, as a rule, considerably simpler than
of the full evolution kernel in general (non-forward) kinematics.

The method suggested in this work is similar to the approach of D.~M\"uller who was the first to use conformal constraints to
determine the form of the renormalized operators to the next-to-leading order (NLO)~\cite{Mueller:1991gd}.
Our technique seems, however, to be better suited for dealing with nonlocal light-ray operators in coordinate representation.
We demonstrated its efficiency by restoring the evolution kernels for twist-two operators in two toy models:
$O(n)$ symmetric $\varphi^4$ theory to the three-loop accuracy and in the matrix $\varphi^3$ model to two loops.

We have calculated the two-loop correction to the operator of  special conformal transformations, $S_+$,
in the $\varphi^4$ theory and observed that it form deviates from the ``natural'' ansatz~(\ref{Conj84}). Thus
the form of a conformal operator suggested in Ref.~\cite{Brodsky:1984xk} does not hold beyond the NLO even in scalar theories.

We expect that the same technique can be applied to gauge theories and in particular to  QCD.
The QCD beta function vanishes for large number of flavors for the critical value of the coupling
$\alpha_s$ in the $d=4-2\epsilon$ dimensions. As a consequence, correlation functions of gauge-invariant operators
are scale-invariant at the critical point. It is believed that QCD correlation functions at the critical point
have to be invariant under conformal transformations as well, although, to our knowledge, this statement
has not been rigorously proven (or disproved) so far.

\begin{acknowledgements}
A.M. is grateful to Dieter M\"uller and Sergey Derkachov for helpful discussions.
This work was supported by the DFG, grant BR2021/5-2.
\end{acknowledgements}

\appendix

\setcounter{equation}{0}
\section{The generator of special conformal transformation
in the $\varphi^4$-theory to the two-loop accuracy}
\label{app:B}
\begin{figure*}
\centerline{ \includegraphics[width=0.85\textwidth]{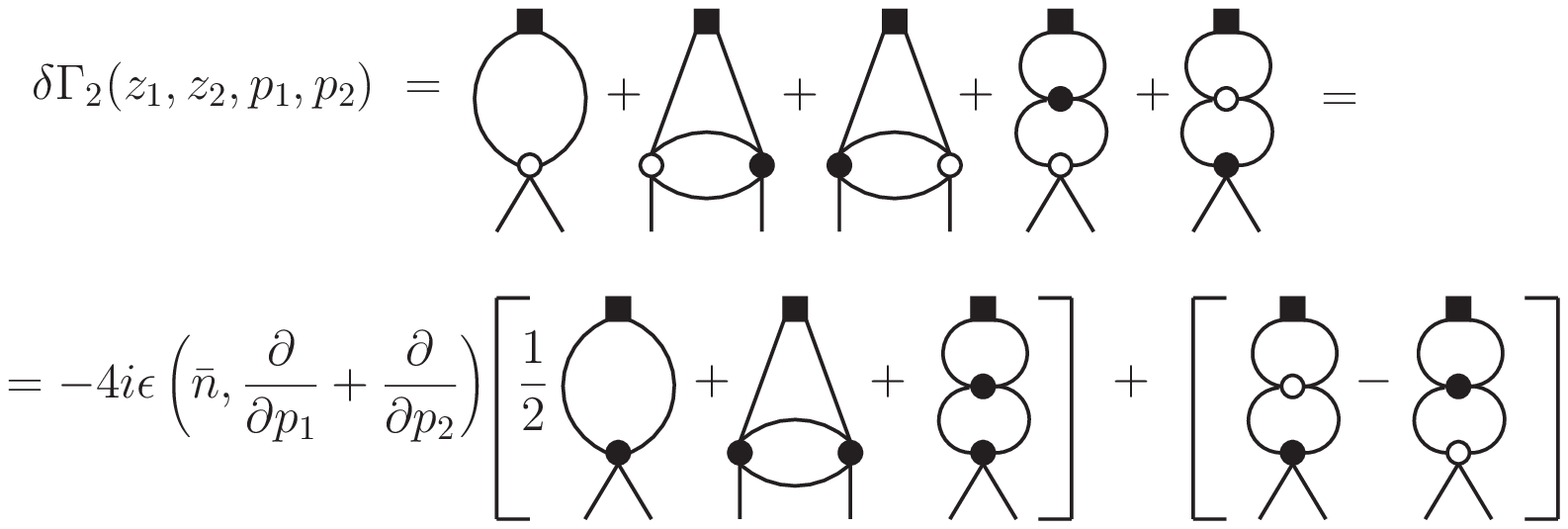}}
\caption{A diagrammatic representation for the 1PI Green function $\delta\Gamma_2(\underline{z},\underline{p})$
(\ref{DeltaSC}) to the two-loop accuracy. The black boxes stand
for the light-ray operator insertion, the filled (black) circles denote the usual $\varphi^4$ interaction vertex
and the open circles correspond to an insertion of $\delta S_R$.}
\label{fig:2}       
\end{figure*}

In  this Appendix we calculate the deformation $\Delta S_+$ of the generator of special conformal transformations
in the $\varphi^4$ theory to the $\mathcal{O}(\epsilon^2)$ accuracy. To this end we need to evaluate
the CWI~(\ref{CWI2}) in the two-loop approximation.

The starting observation is that to the required accuracy
the operator $S_+(u)$ (\ref{S+exp}) turns out to be finite in the $\varphi^4$ theory:
\begin{align}\label{Su1}
S_+(u)=& S_+^{(0)}-\epsilon(z_1+z_2)-\frac{u}{2}[\mathbb{H}^{(1)},(z_1+z_2)]
\notag\\
&
-\frac{u^2}{4}[\mathbb{H}^{(2)},(z_1+z_2)]+\mathcal{O}(u^3)\,.
\end{align}
Indeed, making use of the identities in Eqs.~(\ref{HE2}), (\ref{HwHw})
one can simplify the divergent terms in last two lines in Eq.~(\ref{S+exp}) to
a single term $\sim \mathbb{H}^{(1)}[\mathbb{H}^{(1)},z_1+z_2]$ which happens to be zero
in the $\varphi^4$ theory. This implies, in particular, that the two-point 1PI
Green function $\delta\Gamma_2(\underline{z},\underline{p})$ (\ref{DeltaSC}) is finite
to the two-loop accuracy as well.

A diagrammatic representation for $\delta\Gamma_2(\underline{z},\underline{p})$ is shown in
Fig.~\ref{fig:2}. Taking into account that the one- and two-loop diagrams enter the
expansion with the factors $M^{2\epsilon}$ and $M^{4\epsilon}$, respectively, one gets
\begin{eqnarray}
\delta\Gamma_2(\underline{z},\underline{p})&=&-i(\bar
n,\partial_{p_1}+\partial_{p_2})M\partial_M\Gamma_2(\underline{z},\underline{p})
\nonumber\\&&{}
+\Delta\Gamma_2(\underline{z},\underline{p})+\mathcal{O}(u^3)\,,
\label{twoterms}
\end{eqnarray}
where $\Gamma_2(\underline{z},\underline{p})$ is the usual 1PI Green function (\ref{Gamma2}) and the extra term
$\Delta\Gamma_2(\underline{z},\underline{p})$ stands for the sum of two diagrams in the last square brackets
in the second line in Fig.~\ref{fig:2}.

Using the RG-equation we can rewrite
\begin{align}
M\partial_M\Gamma_2(\underline{z},\underline{p})=-\Big(\beta(u)\partial_u+\mathbb{H}-2\gamma_\varphi\Big)\Gamma_2(\underline{z},\underline{p})\,.
\end{align}
The Green function $\Gamma_2$ on the r.h.s. of this equation can be taken in the one-loop approximation.
Explicit calculation gives
\begin{align}
\Gamma^{(1)}_2(\underline{z},\underline{p})=\Big(1- u \zeta_n\,\mathbb{K}(p_1,p_2)\Big)\Gamma^{(0)}_2(\underline{z},\underline{p})
\end{align}
where, cf. Eq.(\ref{c-coef}),
\begin{align}
     \zeta_n = \{n+2,\,2,\, 0\} = 3 c_n
\end{align}
for the scalar, symmetric traceless and antisymmetric operators, respectively, and
$\Gamma^{(0)}_2(\underline{z},\underline{p})$ is the tree-level 1PI Green function.
Finally, $\mathbb{K}(p_1,p_2)$ is an integral operator that can be written as follows:
\begin{align}
[\mathbb{K}(p_1,p_2) f](z_1,z_2)=
\int_0^1 \!d\alpha\, K(p_1+p_2,\alpha) f(z_{12}^\alpha,z_{12}^\alpha)\,,
\end{align}
with
\begin{align}
K(p,\alpha)=\frac{1}{6\epsilon}
\Big[\Gamma(1+\epsilon)\left(\frac{\alpha\bar\alpha p^2}{4\pi M^2}\right)^{-\epsilon}-1\Big]\,.
\end{align}
Omitting the terms in the $\beta-$function (which vanish at the critical point) one obtains
for the contribution in the first line in Eq.~(\ref{twoterms}):
\begin{eqnarray}\label{BB}
\lefteqn{
-i(\bar n,\partial_{p_1}+\partial_{p_2})M\partial_M\Gamma_2(\underline{z},\underline{p}) =
        }
\nonumber\\&=&
\Big(\mathbb{H}-2\gamma_\varphi\Big)i(\bar
n,\partial_{p_1}+\partial_{p_2})\Gamma^{(1)}_2(\underline{z},\underline{p})
\nonumber\\&=&
\Big(\mathbb{H}-2\gamma_\varphi\Big)\biggl\{-u\zeta_n  \Big[i(\bar
n,\partial_{p_1}+\partial_{p_2})\mathbb{K}(p_1,p_2)\Big]
\nonumber\\&&{}
+ \Big[1-u\zeta_n \mathbb{K}(p_1,p_2)\Big]i(\bar
n,\partial_{p_1}\!+\!\partial_{p_2})\!\biggr\}\Gamma^{(0)}_2(\underline{z},\underline{p})\,.
\end{eqnarray}
Neglecting in this expression terms of higher order than $\mathcal{O}(u^2)$ (note that
in the $\varphi^4$ theory $\gamma_\varphi\sim O(u^2)$) and taking into account that
\begin{align}
i(\bar n,\partial_{p_1}+\partial_{p_2})\Gamma^{(0)}_2(\underline{z},\underline{p})~=~
  -(n\bar n)(z_1+z_2)\Gamma^{(0)}_2(\underline{z},\underline{p})\,
\notag
\end{align}
and
\begin{align}
\mathbb{H}^{(1)}\mathbb{K}(p_1,p_2)(z_1+z_2)~=~\mathbb{H}^{(1)}(z_1+z_2)\mathbb{K}(p_1,p_2)
\notag
\end{align}
one obtains
\begin{eqnarray}\label{BBB}
\lefteqn{
-i(\bar n,\partial_{p_1}+\partial_{p_2})M\partial_M\Gamma_2(\underline{z},\underline{p}) =
        }
\nonumber\\&=&
(n\bar n) \Big[2(z_1+z_2)\gamma_\varphi-\mathbb{H}\,(z_1+z_2)\Big] \Gamma^{(0+1)}_2(\underline{z},\underline{p})
\notag\\&&{}
-u^2\zeta_n \mathbb{H}^{(1)}\Big[i(\bar n,\partial_{p_1}\!+\!\partial_{p_2})\mathbb{K}(p_1,p_2)\Big]
\Gamma^{(0)}_2(\underline{z},\underline{p})\,.
\end{eqnarray}
Using  explicit expression for the one-loop kernel $\mathbb{H}^{(1)}$ (\ref{H1(4)}) the contribution
in the last line in Eq.~(\ref{BBB}) can be simplified to
\begin{align*}\label{}
\Delta' \Gamma_2(\underline{z},\underline{p})=
\frac23 u^2\zeta_n \gamma(\epsilon)\frac{i(\bar n, p_1+p_2)}{(p_1\!+\!p_2)^{2(1+\epsilon)}}\,
\mathbb{H}^{(1)}\Gamma^{(0)}_2(\underline{z},\underline{p})\,,
\end{align*}
where $\gamma(\epsilon)=\Gamma(1+\epsilon)\Gamma^2(1-\epsilon)/\Gamma(2-2\epsilon)$.
It can be checked that this term is canceled by the remaining contribution
$\Delta \Gamma_2(\underline{z},\underline{p})$ in Eq.~(\ref{twoterms}) up to
terms $\mathcal{O}(u^2\epsilon)$:
\begin{align*}
\Delta' \Gamma_2(\underline{z},\underline{p})+\Delta \Gamma_2(\underline{z},\underline{p})= \mathcal{O}(u^2\epsilon).
\end{align*}
Thus one obtains for $\delta \Gamma_2$ at the critical point with the two-loop accuracy
\begin{eqnarray}
\delta\Gamma_2(\underline{z},\underline{p})&=&
(n\bar n) \Big[2(z_1\!+\!z_2)\gamma^*_\varphi
-\mathbb{H}(u_*)(z_1\!+\!z_2)\Big]\Gamma_2(\underline{z},\underline{p})
\nonumber\\&&{}+\mathcal{O}(\epsilon^3)\,.
\end{eqnarray}
Collecting all terms and going over to the coordinate space representation~\footnote{Since
$\gamma_\varphi^*\sim \epsilon^2$ it is sufficient to use the tree level Green function
$\Gamma^{(0)}_2(\underline{z},\underline{x})\sim \delta(x_1-z_1n)\delta(x_2-z_2 n)$ in the
term  $\sim \gamma_\varphi^*\Gamma_2(\underline{z},\underline{x})$.
Hence one can replace in this contribution $\gamma_\varphi^*(z_1n+z_2n,\bar n)\to \gamma_{\varphi}^*(x_1+x_2,\bar n)$
and absorb it in the redefinition of $K_{\bar n}$ generators.} one gets for the CWI~(\ref{CWI2})
\begin{align*}
\Big(K^{x_1}_-(\widetilde \Delta_\varphi)+K^{x_2}_-(\widetilde \Delta_\varphi)
+2(n\bar n) S_+ \Big) \Gamma_2(\underline{z},\underline{x})=\mathcal{O}(\epsilon^3)\,,
\end{align*}
where the $S_+$ generator takes the form~(\ref{S+NL}).

\begin{figure*}
\centerline{ \includegraphics[width=0.75\textwidth]{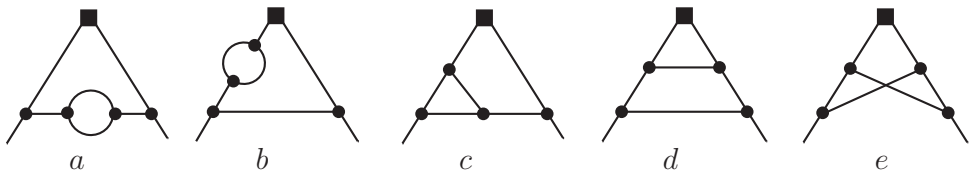}}
\caption{Renormalization of light-ray operators $\mathcal{O}_j(z_1,z_2)$ (filled square) in the $\varphi^3$ theory
to the two-loop accuracy.
}
\label{fig:3}
\end{figure*}


\setcounter{equation}{0}
\section{$su(n)$ projectors}\label{app:C}
A decomposition of the tensor product of two adjoint representations of the $su(n)$ group
contains seven irreducible representations. The projectors onto the scalar and two adjoint
representations have the form
\begin{align}
(P_{1})^{ab}_{a'b'}=&~\frac{1}{n^2-1}\delta^{ab}_{a'b'}\,,
&&
(P_2)^{ab}_{a'b'}=\frac1{n}f^{abc}\, f^{a'b'c}\,,
\notag\\
(P_3)^{ab}_{a'b'}=&~\frac{n}{n^2-4}d^{abc}\,d^{a'b'c}\,.
\end{align}
In addition, we define the projectors $P_4, P_5$ onto symmetric tensors
and the projectors $P_6, P_7$ onto antisymmetric tensors of rank two:
\begin{align}\label{}
P_4=S_+ \Pi_S\,, &&P_5=S_- \Pi_S\,, \notag\\
P_6=A_+ \Pi_A\,, &&P_7=A_- \Pi_A\,,
\end{align}
where
\begin{eqnarray}
\Pi_S&=&\frac12\left(1\!+\!\mathbb{P}\right)\!-\!P_1\!-\!P_3\,,
\quad
\Pi_A~=~\frac12\left(1-\mathbb{P}\right)-P_2\,,
\nonumber\\
S_\pm&=&\frac1{2n}\Big({n\pm 2}\pm n\,\mathbb{R}\Big)\,,
\quad\,
A_{\pm}~=~\frac12\Big(\II\pm i \mathbb{K}\Big),
\end{eqnarray}
$\mathbb{P}$ is the permutation operator,
$\mathbb{P}^{ab}_{a'b'}=\delta^{a}_{b'}\delta^{b}_{a'}$, and
\begin{align*}
\mathbb{R}^{ab}_{a'b'}=d^{aa'c}d^{bb'c}\,, &&
\mathbb{K}^{ab}_{a'b'}=d^{aca'}f^{bcb'}\,.
\end{align*}
The invariant operator $\mathbb{R}$ which arises in the calculation of the one-loop diagram
in Fig.~\ref{fig:1} has the following eigenvalues on the invariant
subspaces, $\mathbb{R}P_j=\lambda_j\, P_j$,
with
\begin{align}\label{lambdas}
&\lambda_1=\frac{n^2-4}{n}\,, &&\lambda_2=\frac{n^2-4}{2n}\,, &&\lambda_3=\frac{n^2-12}{2n}
\notag\\
&\lambda_4=1-\frac{2}n\,,&&\lambda_5=-1-\frac{2}n\,, &&\lambda_{6,7}=-\frac2{n}\,.
\end{align}
For completeness we give the dimensions of the corresponding  subspaces:
\begin{align*}
&\text{dim}V_0=~1\\
&\text{dim}V_{2,3}=~n^2-1,\\
&\text{dim} V_{4(5)}=~n^2(n\pm3)(n\mp 1)/4,\\
&\text{dim} V_{6,7}=~(n^2-1)(n^2-4)/4\,.
\end{align*}
One can check that $\sum_j \lambda_j\text{dim}V_j=0$ that follows from $\tr\mathbb{R}=0$.

\setcounter{equation}{0}
\section{Two-loop evolution kernel in the $\varphi^3$ theory}\label{app:two-loop}

In this  Appendix we collect contributions of individual two-loop diagrams to the renormalization of light-ray
operators $\mathcal{O}_j(z_1,z_2)$ in the $\varphi^3$ theory.
The relevant diagrams are shown in  Fig.~\ref{fig:3} where in all cases we imply that the symmetric diagrams are added.
Let $\Gamma_j^{(a)}$ be a divergent part of the diagram after subtraction of divergent subgraphs.
The results can be presented in the form
\begin{align}
\Gamma_j^{(a)}=u^2\int_0^1d \alpha\int_0^{\bar\alpha} d\beta \chi_j^{(a)}(\alpha,\beta)
\mathcal{O}_j(z_{12}^\alpha,z_{21}^\beta)\,.
\end{align}
We obtain the following expressions:

\noindent
$\bullet~$~Self-energy insertions, Fig.~\ref{fig:3}a,b:
\begin{eqnarray}
\chi_j^{(SE)}(\alpha,\beta)&=&\lambda_j\frac{n^2-4}{24n}\biggl\{\frac3{\epsilon^2}-\frac1\epsilon
\Big[2\ln(1-\alpha-\beta)
\notag\\&&{}
+8-\ln\left(\frac{\bar\alpha\bar\beta}{\alpha\beta}-1\right)\Big]\biggr\}.
\end{eqnarray}

\noindent
$\bullet~$~Vertex correction, Fig.~\ref{fig:3}c:
\begin{eqnarray}
\chi_j^{(V)}(\alpha,\beta)&=&\lambda_j\frac{n^2-12}{8n}\biggl\{-\frac2{\epsilon^2}+\frac1\epsilon
\Big[2\ln(1-\alpha-\beta)
\notag\\&&{}
+6+\ln\left(\frac{\alpha\beta}{\bar\alpha\bar\beta}\right)\Big]\biggr\}.
\end{eqnarray}

\noindent
$\bullet~$~Ladder diagram,  Fig.~\ref{fig:3}d:
\begin{eqnarray}
\chi_j^{(L)}(\alpha,\beta)&=&\frac12\lambda^2_j\biggl\{\left(\frac1{\epsilon^2}+\frac2{\epsilon}\right)
\ln\left(1-\frac{\alpha\beta}{\bar\alpha\bar\beta}\right)
\notag\\&+&
\frac1{2\epsilon}\Big[\ln^2(1\!-\!\alpha\!-\!\beta)-\ln^2\bar\alpha-\ln^2\bar\beta\Big]\biggr\}.
\end{eqnarray}

\noindent
$\bullet~$ Crossed diagram, Fig.~\ref{fig:3}e:
\begin{equation}
\chi_j^{(C)}(\alpha,\beta)=-\nu_j\frac1{2\epsilon}\ln\left(1-\frac{\alpha\beta}{\bar\alpha\bar\beta}\right).
\end{equation}
Here
\begin{align}\label{nus}
&\nu_2=\nu_6=\nu_7=0\,,
\notag\\
&\nu_1= {(n^2-4)(n^2-12)}/{(2n^2)}\,,
\notag\\
&\nu_3=- {4(n^2-10)}/{n^2}\,,
\notag\\
&\nu_{4(5)}= \pm{(n\mp2)(n^2\pm4n-8)}/{(2n^2)}\,.
\end{align}


\end{document}